\documentclass[superscriptaddress,twocolumn,showpacs,
amssymb,amsmath,nobibnotes,aps,prd,
nofootinbib]{revtex4}
\pdfoutput=1
\usepackage{graphicx,subfigure,bm,color,psfrag,hyperref}
\usepackage{amsfonts}
\usepackage{lipsum}
\usepackage{mathtools}
\usepackage{verbatim}
\usepackage{color}

\begin{document}

\title{Interacting scenarios with dynamical dark energy: observational constraints and alleviation of the $H_0$ tension}

\author{Supriya Pan}
\email{supriya.maths@presiuniv.ac.in}
\affiliation{Department of Mathematics, Presidency University, 86/1 College Street, 
Kolkata 700073, India.}

\author{Weiqiang Yang}
\email{d11102004@163.com}
\affiliation{Department of Physics, Liaoning Normal University, Dalian, 116029, P. R. 
China.}

\author{Eleonora Di Valentino}
\email{eleonora.divalentino@manchester.ac.uk}
\affiliation{Jodrell Bank Center for Astrophysics, School of Physics and Astronomy, 
University of Manchester, Oxford Road, Manchester, M13 9PL, UK.}

\author{Emmanuel N. Saridakis}
\email{msaridak@phys.uoa.gr}
\affiliation{Department of Physics, National Technical University of Athens, Zografou 
Campus GR 157 73, Athens, Greece}
\affiliation{Department of Astronomy, School of Physical Sciences, University of Science 
and Technology of China, Hefei 230026, P.R. China}
\affiliation{Chongqing University of Posts \& Telecommunications, Chongqing, 400065, P.R. 
China}

\author{Subenoy Chakraborty}
\email{schakraborty.math@gmail.com}
\affiliation{Department of Mathematics, Jadavpur University, Kolkata 700032, West Bengal, 
India}

\begin{abstract}
We investigate interacting scenarios which belong to a wider class, since they include a
dynamical dark energy component whose equation of state follows various  one-parameter 
parametrizations. We confront them with the latest observational data from Cosmic 
Microwave Background (CMB), Joint light-curve (JLA) sample from Supernovae Type Ia, 
Baryon Acoustic Oscillations (BAO), Hubble parameter measurements from Cosmic 
Chronometers (CC)  and a gaussian prior on the Hubble parameter $H_0$. In all 
examined scenarios we find a non-zero interaction,  nevertheless the non-interacting case 
is allowed within 2$\sigma$.
Concerning the current value of the dark energy equation of state for all 
combination of datasets it always lies in the phantom regime at more than two/three 
standard deviations. Finally, for all interacting models, independently of 
the combination of datasets considered, the estimated values of the present Hubble 
parameter $H_0$ are greater compared to the $\Lambda$CDM-based Planck's estimation and 
close to the local measurements, thus  alleviating the $H_0$ tension.

\end{abstract}


\maketitle
\section{Introduction}

 After almost 20 years from the detection of late-time universe  acceleration, and due to 
the appearance of a huge amount of data, people are still looking for the actual 
underlying theory that could explain it.   In general, there are two widely accepted 
approaches that could accommodate it. The first one is the introduction of some 
hypothetical dark energy fluid \cite{Copeland:2006wr} in the context of Einstein's 
gravitational theory. The second one is to consider   modified or alternative 
gravitational theories where the extra geometrical terms may reproduce the effects of 
dark energy \cite{Capozziello:2011et,Cai:2015emx,Nojiri:2017ncd}.  

On the other hand,  a mutual interaction between the dark matter and dark energy sectors  
was initially introduced in order to 
investigate the cosmological constant problem \cite{Wetterich-ide1}. However, later on 
it was found that it could also alleviate the cosmic coincidence problem in a natural way 
\cite{Amendola-ide1,Amendola-ide2,Cai:2004dk,Pavon:2005yx,delCampo:2008sr,delCampo:2008jx}
, and this 
led to  many investigations of interacting cosmology
\cite{Barrow:2006hia,Amendola:2006dg,He:2008tn,Chen:2008ft,Basilakos:2008ae,Gavela:2009cy,Sadjadi:2009sp,
Jamil:2009eb,Chen:2011cy,Pan:2013rha,Yang:2014vza,Yang:2014gza,Nunes:2014qoa,
Faraoni:2014vra,Salvatelli:2014zta,Yang:2014hea,Pan:2012ki,Li:2015vla,Nunes:2016dlj,
Yang:2016evp,Kumar:2016zpg,Pan:2016ngu,Mukherjee:2016shl,Erdem:2016hqw,Sharov:2017iue,
Shahalam:2017fqt,Guo:2017hea,Cai:2017yww,Yang:2017yme,Santos:2017bqm,Yang:2017ccc,
vandeBruck:2017idm,Pan:2017ent,Xu:2017rfo,Yang:2018xlt,Yang:2018ubt,Yang:2018pej,
Cardenas:2018qcg,Odintsov:2018awm,vonMarttens:2018iav,Yang:2018qec,Bonici:2018qli,
Asghari:2019qld,Paliathanasis:2019hbi,Pan:2019jqh,Feng:2019mym,Li:2019loh,Yang:2019bpr,
Yang:2019vni,Oikonomou:2019nmm} (also see \cite{Bolotin:2013jpa,Wang:2016lxa} for recent 
reviews on interacting dark matter-dark energy scenarios).
An additional advantage of interacting scenarios is the easy realization of the 
phantom-divide crossing without theoretical ambiguities 
\cite{Wang:2005jx,Sadjadi:2006qb,Pan:2014afa}. Finally, interacting scenarios prove to be 
efficient 
in alleviating the two known tensions of modern cosmology, namely the  $H_0$ 
one~\cite{DiValentino:2015ola,DiValentino:2016hlg,Kumar:2017dnp,DiValentino:2017iww,
DiValentino:2017zyq,Renk:2017rzu,
DiValentino:2017gzb,DiValentino:2017oaw,Fernandez-Arenas:2017isq,DiValentino:2017rcr,
Khosravi:2017hfi,Mortsell:2018mfj,Yang:2018euj,DEramo:2018vss,Yang:2018uae,Guo:2018ans,
Yang:2018qmz,
Poulin:2018cxd,Zhang:2018air,Kreisch:2019yzn,Martinelli:2019dau,Pandey:2019plg,
Vattis:2019efj,
Kumar:2019wfs,Agrawal:2019lmo,Yang:2019qza,Yang:2019uzo,DiValentino:2019exe,Yang:2019nhz}, and the $\sigma_8$ one 
\cite{Pourtsidou:2016ico,An:2017crg,DiValentino:2018gcu,Kumar:2019wfs}.

Despite the extended investigation of interacting scenarios,  the choice of the  
interaction function remains unknown. Thus, in general one considers   phenomenological 
models for the interaction form and explores the cosmological dynamics confronting with 
observational data. The complication in the above procedure, which is not usually taken 
into account, is that in principle apart from the  unknown interaction form one has also 
the ambiguity in the dark-energy equation-of-state parameter. Hence, in the present work 
we are interested in performing a systematic confrontation of interacting dark energy 
scenarios, considering however all well-studied parametrizations for the  dark-energy 
equation-of-state parameter. Only such a complete and consistent analysis can extract 
safe results about the observational validity of the examined scenarios.

We consider interacting scenarios in which the dark energy equation of state is 
parametrized with forms that include one free parameter. Such one-parameter models are more economical comparing to the two-parameter ones, and 
moreover recently it was found that these one-parameter 
dynamical dark-energy parametrizations  are very efficient   in alleviating the
$H_0$ tension in the simple non-interacting framework \cite{Yang:2018qmz}. 
This motivates us to consider a wider picture in which the 
interaction should be allowed too, and to check whether the $H_0$ tension is still 
released, since it has been argued that an allowance of a non-gravitational interaction 
between 
dark  matter and dark energy naturally increases the error bars on $H_0$ (due to the 
existing correlation between $H_0$ and the coupling parameter of the interaction models) 
and consequently  alleviates the corresponding tension  
\cite{Kumar:2017dnp,DiValentino:2017iww,Yang:2018euj,Yang:2018uae}.   Thus, essentially 
the present work aims to investigate whether the release of $H_0$ tension discussed in 
\cite{Yang:2018qmz} is   influenced by the presence of an 
interaction between dark matter and dark energy.

The work has been organized in the following manner. In section \ref{sec-2} we describe 
the  basic equations for any interacting dark energy model at the background and 
perturbative levels. Additionally, we present various  one-parameter $w_x$ 
parametrizations. Section \ref{sec-data} deals with the observational 
data that we   consider in this work. In section \ref{sec-results} we 
describe the main observational results extracted for all the examined scenarios. 
Moreover, in section \ref{sec-bayesian} we compute the  Bayesian evidences of the models 
with respect to the reference  $\Lambda$CDM  paradigm. 
Finally, in section \ref{sec-summary} we conclude the present work with a brief summary 
of  all findings.

\section{Cosmological equations in  interacting scenarios}
\label{sec-2}

The  universe is well described by the homogeneous and 
isotropic 
Friedman-Lema\^{i}tre-Robertson-Walker (FLRW) line element given by 
\begin{eqnarray}
ds^2 = -dt^2 + a^2 (t) \left[\frac{dr^2}{1-Kr^2} + \left( d\theta ^2 + \sin^2 \theta 
d\phi^2\right)  \right],
\end{eqnarray}
where $a(t)$ is the expansion scale factor and $K=0, -1, +1$ corresponds 
respectively to   flat, open and closed spatial geometry. Since   observations
imply almost spatial flatness,  we shall restrict 
ourselves to $K= 0$ throughout the work.  

The total matter content of the universe constitutes  to radiation, baryons, pressure-less dark matter and 
a dark energy fluid (that may be real fluid or an effective one arising from modified 
gravity). Moreover, we allow  the dark matter and dark energy to have a mutual 
(non-gravitational) interaction, while the remaining two fluids follow the usual 
conservation laws. Hence, the   Friedmann equation is given by 
\begin{eqnarray}\label{Hubble}
H^2 = \frac{8 \pi G}{3} \left( \rho_r + \rho_b + \rho_c + \rho_x  \right),
\end{eqnarray}
in which $H \equiv \dot{a}/a$ is the Hubble rate, and $\rho_i$ is the 
energy density of the $i$-th fluid sector (with  $ i=r$ for radiation, $ i = b$ 
for baryons, $i = c $ for cold or pressure-less dark matter and $i =x$ for dark energy).  
The conservation equation of the total fluid $\rho_{\rm tot} = \rho_r +\rho_b +\rho_c 
+\rho_x$, is given by 
\begin{eqnarray}\label{cons-tot}
\dot{\rho}_{\rm tot} + 3 H \left( \rho_{\rm tot} + p_{\rm tot} \right) = 0,
\end{eqnarray}
where $p_{\rm tot}$ is the total pressure of the fluids defined as $p_{\rm tot} = p_r 
+p_b 
+p_c +p_x$. Since radiation and baryons satisfy their own conservation equations, namely, 
$\dot{\rho}_b + 3 H \rho_b =0$ and  $\dot{\rho}_r + 4 H \rho_r =0$,  then  the 
conservation equation for the total fluid (\ref{cons-tot}) gives rise to 
\begin{eqnarray}\label{cons-dark-sector}
\dot{\rho}_{\rm Dark} + 3 H \left(p_{\rm Dark }+ \rho_{\rm Dark}\right) =0,
\end{eqnarray} 
where $\rho_{\rm Dark} = \rho_c +\rho_x$ and $p_{\rm Dark} = p_c +p_x$.

In interacting cosmology one  splits 
  the conservation equation for the dark sector (\ref{cons-dark-sector})  into 
\begin{eqnarray}\label{cons-dm}
\dot{\rho}_c + 3 H \rho_c = -Q (t),
\end{eqnarray} 
and 
\begin{eqnarray}\label{cons-de}
\dot{\rho}_x + 3 H (1+w_x) \rho_x = Q (t),
\end{eqnarray} 
by introducing a new function $Q (t)$, that actually characterizes the rate of energy 
transfer between these dark fluids. Thus, whenever the interaction $Q$ is prescribed, 
using the conservation equations (\ref{cons-dm}), (\ref{cons-de}) as well as the 
Friedmann equation (\ref{Hubble}), one can determine the dynamics of this interacting 
scenario. 

Since the nature of both dark fluids is unknown, there is an ambiguity in the choice of 
the interaction function. Thus, in general one 
considers phenomenological choices for $Q$, and through observational confrontation 
results to the best interaction model.  In the present work we will focus on a well 
  motivated interaction  that induces stable perturbations \cite{Yang:2017zjs}:     
\begin{eqnarray}\label{Q}
Q = 3\xi H  (1+w_x) \rho_x,
\end{eqnarray} 
with $\xi $  the coupling parameter   characterizing the interaction strength. 

Let us  briefly describe the perturbation equations for an interacting dark 
energy  model following \cite{Mukhanov, Ma:1995ey, Malik:2008im}. 
 The scalar perturbations of the FLRW metric read as
\begin{eqnarray}
ds^{2}=a^{2}(\tau )\Bigg[-(1+2\phi )d\tau ^{2}+2\partial _{i}Bd\tau dx^{i}\notag\\+
\Bigl((1-2\psi )\delta _{ij}+2\partial _{i}\partial _{j}E\Bigr)dx^{i}dx^{j}%
\Bigg],
\end{eqnarray}%
where $\tau$ is the conformal time and the  $\phi $, $B$, $\psi $ and  $E$ are the 
gauge-dependent scalar perturbation quantities. 
Additionally, for an interacting universe the conservation equations become
\cite{Majerotto:2009np, Valiviita:2008iv, Clemson:2011an}
\begin{equation}
\nabla _{\nu }T_{A}^{\mu \nu }=Q_{A}^{\mu },~~~~\sum\limits_{\mathrm{A}}{%
Q_{A}^{\mu }}=0,
\end{equation}
where $A$ is used to represent either pressure-less dark matter (then $A =c$) or dark 
energy (then $A=x$). 
Here, the quantity $Q_{A}^{\mu }$ takes the following expression   
\begin{eqnarray}
Q_{A}^{\mu }=(Q_{A}+\delta Q_{A})u^{\mu }+a^{-1}(0,\partial
^{i}f_{A}), 
\end{eqnarray}
relative to the four-velocity $u^{\mu }$, in which $Q_A$ presents the background energy
transfer (i.e. $Q_A = Q$) and $f_A$ is the momentum transfer potential.
We restrict ourselves to the simplest possibility 
 following the earlier works  
\cite{Majerotto:2009np, Valiviita:2008iv, Clemson:2011an}, i.e. we assume that the 
momentum transfer potential is 
zero in the rest frame of the dark matter, from which 
one can derive that
$k^{2}f_{A}=Q_{A}(\theta -\theta _{c})$ (here $k$ is the wave number;
$\theta = \theta_{\mu}^{\mu}$ is the volume expansion scalar of the total fluid, and
$\theta_c$ is the the volume expansion scalar for the CDM fluid). 

We proceed by applying  the synchronous gauge to derive the perturbation 
equations
for the interacting scenarios.   Thus, in the synchronous gauge we have
$\phi =B=0$, $\psi =\eta $, and $k^{2}E=-h/2-3\eta $ ($h$ and $\eta$ are the metric 
perturbations, see \cite{Ma:1995ey} for details). Additionally, we assume the absence 
of an anisotropic stress, and we define the density perturbations for the 
fluid  $A$   by  $\delta _{A}=\delta \rho _{A}/\rho_{A}$. The resulting perturbation 
equations become
\begin{eqnarray}
\delta _{x}^{\prime } &=&-(1+w_{x})\left( \theta _{x}+\frac{h^{\prime }}{2}%
\right)-3\mathcal{H}w_{x}^{\prime }\frac{%
\theta _{x}}{k^{2}} 
\nonumber\\
&&-3\mathcal{H}(c_{sx}^{2}-w_{x})\left[ \delta _{x}+3\mathcal{H}%
(1+w_{x})\frac{\theta _{x}}{k^{2}}\right] 
\notag \\
&&+\frac{aQ}{\rho _{x}}\left[ -\delta _{x}+\frac{\delta Q}{Q}+3\mathcal{H}%
(c_{sx}^{2}-w_{x})\frac{\theta _{x}}{k^{2}}\right] , 
\label{eq:perturbation1}\\
\theta _{x}^{\prime } &=&-\mathcal{H}(1-3c_{sx}^{2})\theta _{x}+\frac{%
c_{sx}^{2}}{(1+w_{x})}k^{2}\delta _{x}\nonumber\\
&&
+\frac{aQ}{\rho _{x}}\left[ \frac{%
\theta _{c}-(1+c_{sx}^{2})\theta _{x}}{1+w_{x}}\right] , \label{eq:perturbation2}\\
\delta _{c}^{\prime } &=&-\left( \theta _{c}+\frac{h^{\prime }}{2}\right) +%
\frac{aQ}{\rho _{c}}\left( \delta _{c}-\frac{\delta Q}{Q}\right) , 
\label{eq:perturbation3}\\
\theta _{c}^{\prime } &=&-\mathcal{H}\theta _{c},  \label{eq:perturbation4}
\end{eqnarray}
where $\mathcal{H}=aH$  is the conformal Hubble rate and 
in (\ref{eq:perturbation1}), (\ref{eq:perturbation2}),
(\ref{eq:perturbation3}) the factor 
$\delta Q/Q$ incorporates the perturbations for the
Hubble rate $\delta H$. We mention that using   $\delta H$ one 
can easily find the gauge invariant perturbation equations   \cite{Gavela:2009cy}.

We close this section by introducing the $w_x$ parametrizations having only one free 
parameter $w_0$, namely the present value of the dark energy equation of state  
\cite{Yang:2018qmz}:
\begin{eqnarray}
&&{\rm Model~I}:~~w_x(a)=w_0a[1-\log(a)],\label{model1}\\
&&{\rm Model~II}:~~w_x (a) = w_0 a \exp(1-a), \label{model2}\\
&&{\rm Model~III}:~~w_x(a)=w_0a[1+\sin(1-a)],\label{model3}\\
&&{\rm Model~IV}:~~w_x(a)=w_0a[1+\arcsin(1-a)].\label{model4}
\end{eqnarray}

Thus, in summary we consider the interaction model (\ref{Q}) with four different dark 
energy equations of state given in (\ref{model1})-(\ref{model4}). From now on we   
identify the interaction model (\ref{Q}) with  $w_x$ of (\ref{model1}) as IDE1,
interaction model (\ref{Q}) with    (\ref{model2}) as IDE2,
interaction model (\ref{Q}) with  (\ref{model3}) as IDE3, and finally the  
interaction model (\ref{Q}) with    (\ref{model4}) as IDE4.

\section{Observational data}
\label{sec-data}

In this section we describe the observational data that we   use to investigate the 
interacting dark energy models, and we provide  a brief description on the methodology.

\begin{itemize}

\item The data from cosmic microwave background (CMB) observations are very powerful to 
analyze the cosmological models. Here we use the high-$\ell$ temperature and 
polarization as well as the low-$\ell$ temperature and polarization 2015 CMB angular 
power 
spectra from the Planck experiment (Planck TT, TE, EE + lowTEB)  \cite{Adam:2015rua, 
Aghanim:2015xee}. 

\item We include the  Joint light-curve analysis (JLA) sample from Supernovae Type Ia 
data 
\cite{Betoule:2014frx}. 

\item We use the Baryon acoustic oscillations (BAO) distance measurements from the 
following references \cite{Beutler:2011hx, Ross:2014qpa,Gil-Marin:2015nqa}. 

\item We consider the measurements of the Hubble parameter at various redshifts from the 
Cosmic Chronometers (CC)  \cite{Moresco:2016mzx}.

\item We adopt a gaussian prior on the Hubble constant (R19) $H_0=74.02\pm1.42$ as 
obtained from SH0ES~\cite{Riess:2019cxk}.

\end{itemize}

\begin{table}[ht]
\begin{center}
\begin{tabular}{c|c}
Parameter                    & Prior \\
\hline 
$\Omega_{b} h^2$             & $[0.005,0.1]$\\
$\Omega_{c} h^2$             & $[0.01,0.99]$\\
$\tau$                       & $[0.01,0.8]$\\
$n_s$                        & $[0.5, 1.5]$\\
$\log[10^{10}A_{s}]$         & $[2.4,4]$\\
$100\theta_{MC}$             & $[0.5,10]$\\
$w_0$                        & $[-2, 0]$\\  
$\xi$                        & $[0, 2]$\\
\end{tabular}
\end{center}
\caption{The table shows the priors imposed on various free parameters of the interacting 
scenarios during the statistical analysis. }
\label{tab:priors}
\end{table}

\squeezetable 
\begingroup           
\begin{center}                                                                            
\begin{table*} [!]                                                                        
 
\scalebox{0.9}
{\begin{tabular}{cccccccccccccccc}          
\hline\hline

Parameters & CMB & CMB+BAO & CMB+BAO+JLA & CMB+BAO+JLA+CC & CMB+R19 & CMB+BAO+R19 \\ 
\hline
$\Omega_c h^2$ & $    0.1209_{-    0.0020-    0.0035}^{+    0.0015+    0.0038}$ &  $    
0.1194_{-    0.0013-    0.0025}^{+    0.0013+    0.0025}$ & $    0.1187_{-    0.0012-    
0.0022}^{+    0.0012+    0.0023}$ & $    0.1187_{-    0.0012-    0.0025}^{+    0.0012+    
0.0025}$  & $    0.1202_{-    0.0016-    0.0027}^{+    0.0015+    0.0028}$  & $    
0.1198_{-    0.0012-    0.0024}^{+    0.0012+    0.0024}$   \\

$\Omega_b h^2$ & $    0.02220_{-    0.00015-    0.00031}^{+    0.00016+    0.00031}$ & $  0.02223_{-    0.00015-    0.00029}^{+    0.00016+    0.00028}$ & $    0.02226_{-    
0.00014-    0.00027}^{+    0.00014+    0.00028}$ & $    0.02226_{-    0.00014-    
0.00028}^{+    0.00014+    0.00028}$ & $    0.02219_{-    0.00015-    0.00028}^{+    
0.00015+    0.00028}$ & $    0.02222_{-    0.00016-    0.00031}^{+    0.00015+    
0.00030}$ \\

$100\theta_{MC}$ & $    1.04036_{-    0.00031-    0.00070}^{+    0.00037+    0.00065}$ & 
$ 
   1.04049_{-    0.00032-    0.00065}^{+    0.00032+    0.00063}$ & $    1.04058_{-    
0.00030-    0.00058}^{+    0.00030+    0.00060}$ & $    1.04059_{-    0.00031-    
0.00060}^{+    0.00029+    0.00059}$ & $    1.04040_{-    0.00032-    0.00060}^{+    
0.00035+    0.00056}$  & $    1.04044_{-    0.00033-    0.00060}^{+    0.00033+    
0.00062}$  \\

$\tau$ & $    0.080_{-    0.018-    0.035}^{+    0.018+    0.035}$ & $    0.086_{-    
0.018-    0.036}^{+    0.018+    0.036}$  & $    0.092_{-    0.018-    0.034}^{+    
0.017+ 
   0.034}$ & $    0.092_{-    0.017-    0.034}^{+    0.018+    0.033}$  & $    0.081_{-   
 0.019-    0.032}^{+    0.017+    0.034}$ & $    0.082_{-    0.018-    0.037}^{+    0.018+ 0.037}$  \\

$n_s$ & $    0.9734_{-    0.0044-    0.0082}^{+    0.0044+    0.0081}$ & $    0.9748_{-  0.0041-    0.0082}^{+    0.0042+    0.0084}$ & $    0.9764_{-    0.0040-    0.0079}^{+    
0.0041+    0.0081}$ &  $    0.9766_{-    0.0042-    0.0083}^{+    0.0041+    0.0088}$  & 
$ 0.9734_{-    0.0047-    0.0083}^{+    0.0047+    0.0082}$ & $    0.9739_{-    0.0041-   0.0077}^{+    0.0040+    0.0083}$  \\

${\rm{ln}}(10^{10} A_s)$ & $    3.103_{-    0.034-    0.067}^{+    0.034+    0.067}$ & $  3.114_{-    0.034-    0.069}^{+    0.037+    0.070}$ & $    3.124_{-    0.035-    
0.065}^{+    0.034+    0.067}$ & $    3.124_{-    0.033-    0.065}^{+    0.036+    
0.065}$  & $    3.107_{-    0.037-    0.068}^{+    0.033+    0.068}$ & $    3.108_{-    0.034-    
0.072}^{+    0.035+    0.072}$ \\

$w_0$ & $  <-1.30\,<-1.07$ & $   -1.207_{-    0.054-    0.11}^{+    0.058+    0.11}$ & $  
 -1.137_{-    0.037-    0.073}^{+    0.037+    0.074}$ & $   -1.138_{-    0.040-    
0.085}^{+    0.041+    0.082}$ & $   -1.306_{-    0.046-    0.10}^{+    0.053+    0.09}$  
& $   -1.263_{-    0.040-    0.088}^{+    0.041+    0.089}$  \\

$\xi$ & $ <0.0068\,<0.014$ & $ <0.0047\,<0.0071$ & $    0.0030_{-    0.0020}^{+    
0.0019}\,<0.0058$ & $    0.0031_{-    0.0021}^{+    0.0019}\,<0.0059$  & $    <0.0061\, <0.0088$ & $    <0.0054\, <0.0080$ \\

$\Omega_{m0}$ & $    0.231_{-    0.081-    0.098}^{+    0.077+    0.109}$ & $    0.283_{- 
 0.011-    0.022}^{+    0.011+    0.022}$ & $    0.297_{-    0.0088-    0.0169}^{+    
0.0087+    0.0171}$ & $    0.296_{-    0.0086-    0.0176}^{+    0.0088+    0.0175}$ & $   
 0.2632_{-    0.0095-    0.020}^{+    0.0095+    0.020}$ & $    0.271_{-    0.0074-    
0.016}^{+    0.0077+    0.016}$ \\

$\sigma_8$ & $    0.932_{-    0.090-    0.143}^{+    0.087+    0.144}$ & $    0.855_{-    
0.021-    0.038}^{+    0.021+    0.040}$ & $    0.835_{-    0.017-    0.034}^{+    0.018+ 
 0.034}$ & $    0.836_{-    0.017-    0.033}^{+    0.017+    0.035}$ & $    0.882_{-    
0.020-    0.033}^{+    0.018+    0.034}$ & $    0.870_{-    0.017-    0.035}^{+    0.019+ 
 0.032}$ \\

$H_0$ & $   81_{-   14-   16}^{+   13+   17}$ & $   71.0_{-    1.5-    3.0}^{+    1.5+    
2.9}$ & $   69.1_{-    1.0-    2.0}^{+ 1.0 +  2.0}$ & $   69.2_{-    1.1-    2.1}^{+    
1.0+    2.1}$  & $   73.8_{-    1.5-    2.6}^{+    1.2+    3.0}$  & $   72.6_{-    1.0-   
 2.0}^{+    1.0+    2.2}$ \\

$S_8$ & $   0.820_{- 0.045 - 0.08}^{+ 0.038  + 0.061}$ & $   0.824_{-    0.017-    
0.055}^{+    0.019+    0.035}$ & $   0.823_{-    0.017-    0.057}^{+ 0.020 +  0.036}$ & 
$  0.825_{-    0.015-    0.051}^{+    0.017+    0.034}$  & 
$   0.826_{-    0.016-    0.030}^{+    0.026+    0.031}$  & $   0.826_{-    0.016-    
0.030}^{+    0.015+    0.030}$\\

\hline\hline                                                            
\end{tabular}     
}  
\caption{68\% and 95\% 
confidence-level constraints on the interacting scenario IDE1 with the dark energy 
equation of state $w_x(a)=w_0a[1-\log(a)]$ (Model I) for various observational datasets. 
Here $\Omega_{m0}$ is the present value of the total matter density 
parameter $\Omega_m = \Omega_{b}+\Omega_{c}$, and $H_0$ is in   units of km/sec/Mpc. }
\label{tab:modelI}                  
\end{table*}                         
\end{center}                
\endgroup

\begin{figure*}[!]
\includegraphics[width=0.72\textwidth]{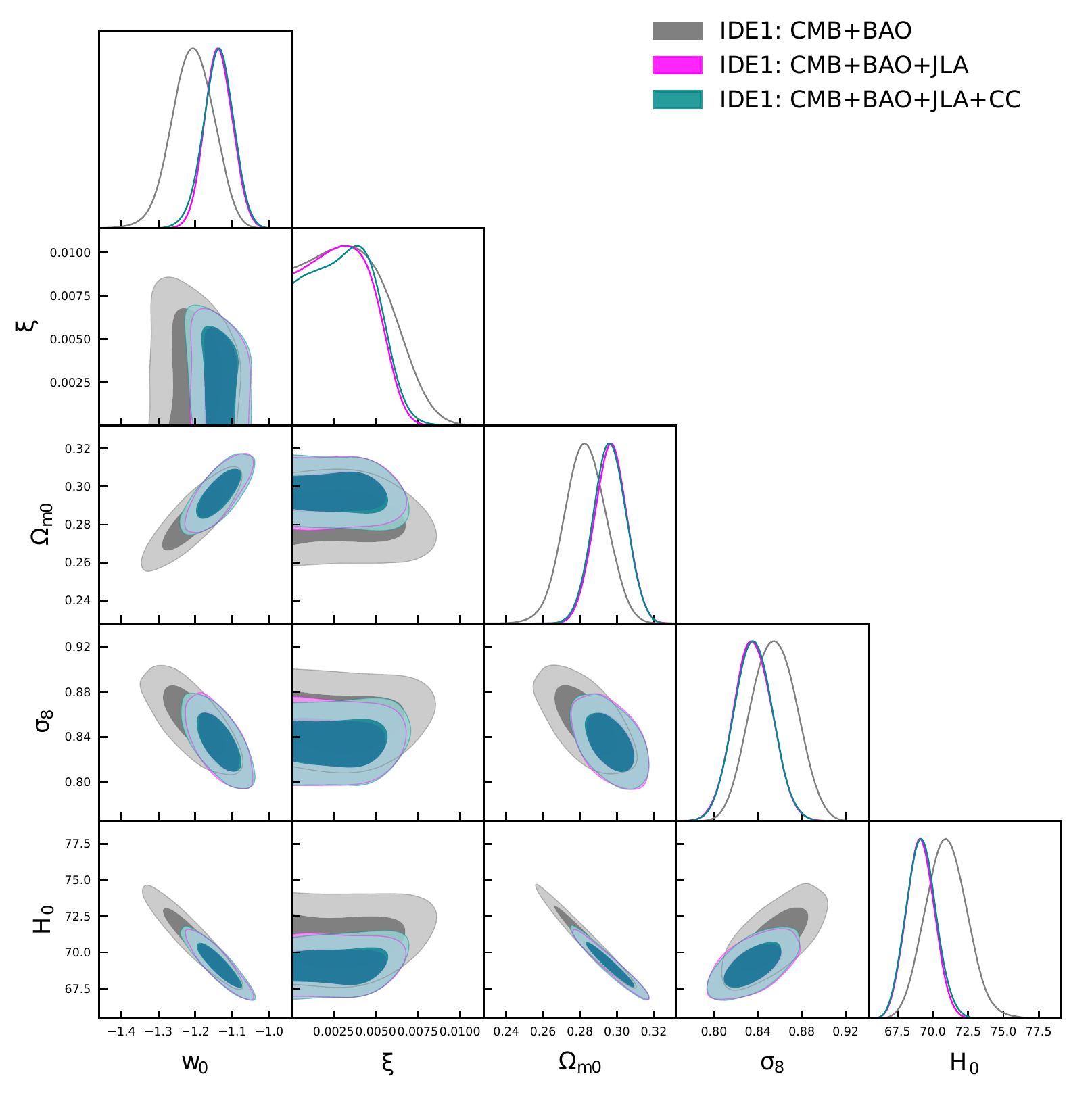}
\caption{{\it{The 68\% and 95\% Confidence Level (CL) contour plots between various 
combinations of the model parameters of scenario IDE1, using different observational 
astronomical datasets. Additionally we  display the one-dimensional marginalized 
posterior distributions of some free parameters.}}}
\label{fig-contour-ide1}
\end{figure*}   

\begin{figure*}
\includegraphics[width=0.68\textwidth]{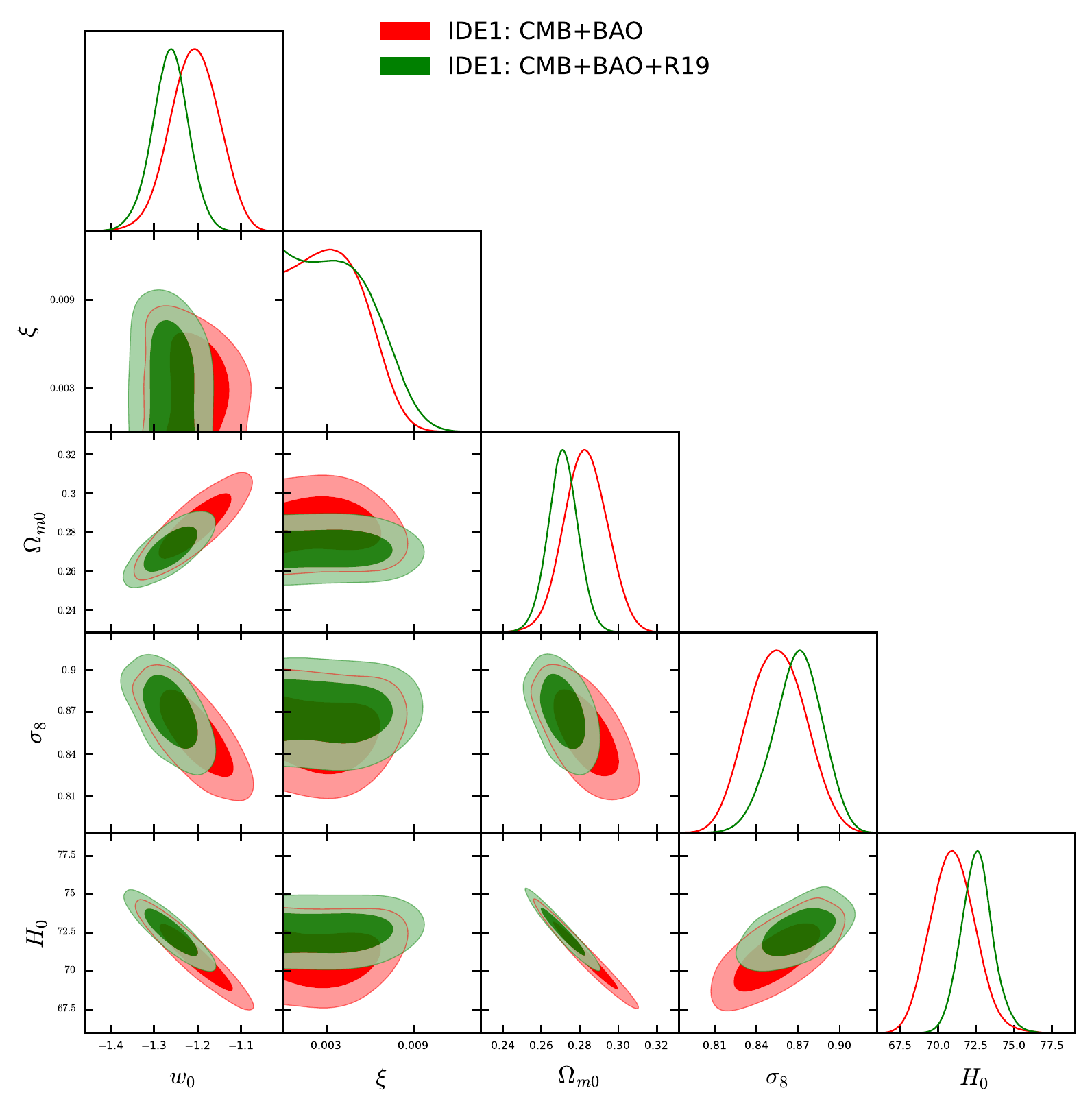}
\caption{
{\it{The 68\% and 95\% CL contour plots between various combinations of the 
model parameters of scenario IDE1 using only   the CMB+BAO 
and CMB+BAO+R19 datasets, and the corresponding
one-dimensional 
marginalized posterior distributions.}} }
\label{fig-contour-ide1-R19}
\end{figure*}

In order  to extract the observational constraints on the model parameters of the 
interaction scenarios, we use the efficient cosmological code \texttt{cosmomc} 
\cite{Lewis:2002ah, Lewis:1999bs}, a markov chain monte carlo package which (i) has a 
convergence diagnostic and (ii) supports the Planck 2015 likelihood code 
\cite{Aghanim:2015xee}.  The dimension of the parameters space for all 
  interaction scenarios is eight, where  
$\mathcal{P} \equiv\Bigl\{\Omega_bh^2, \Omega_{c}h^2, 100\theta_{MC}, \tau, w_0, \xi, 
n_s, 
log[10^{10}A_S]\Bigr\}.$ Here $\Omega_bh^2$ is the physical baryon density, 
$\Omega_{c}h^2$ is the physical density for cold dark matter, 
$100 \theta_{MC}$ denotes the ratio of the sound horizon to the angular diameter 
distance, 
$\tau$ denotes the reionization optical depth, $n_s$ is the scalar spectral index, $A_S$ 
is the amplitude of the primordial scalar power spectrum, $w_0$  is the current value of 
the dark energy parameter, and $\xi$ is the coupling parameter of the interaction. In 
Table \ref{tab:priors} we summarize the flat priors on the model parameters during the 
statistical analysis.

\section{Results and implications}
\label{sec-results}

In this section we extract the  observational constraints   on the present four 
interacting dark energy scenarios where dark energy has a time-dependent 
equation-of-state 
parameter. For all   interacting scenarios we have performed several analyses using the 
observational data described in section \ref{sec-data}. 

\begin{figure*}
\begin{center}
\includegraphics[width=1.07\textwidth]{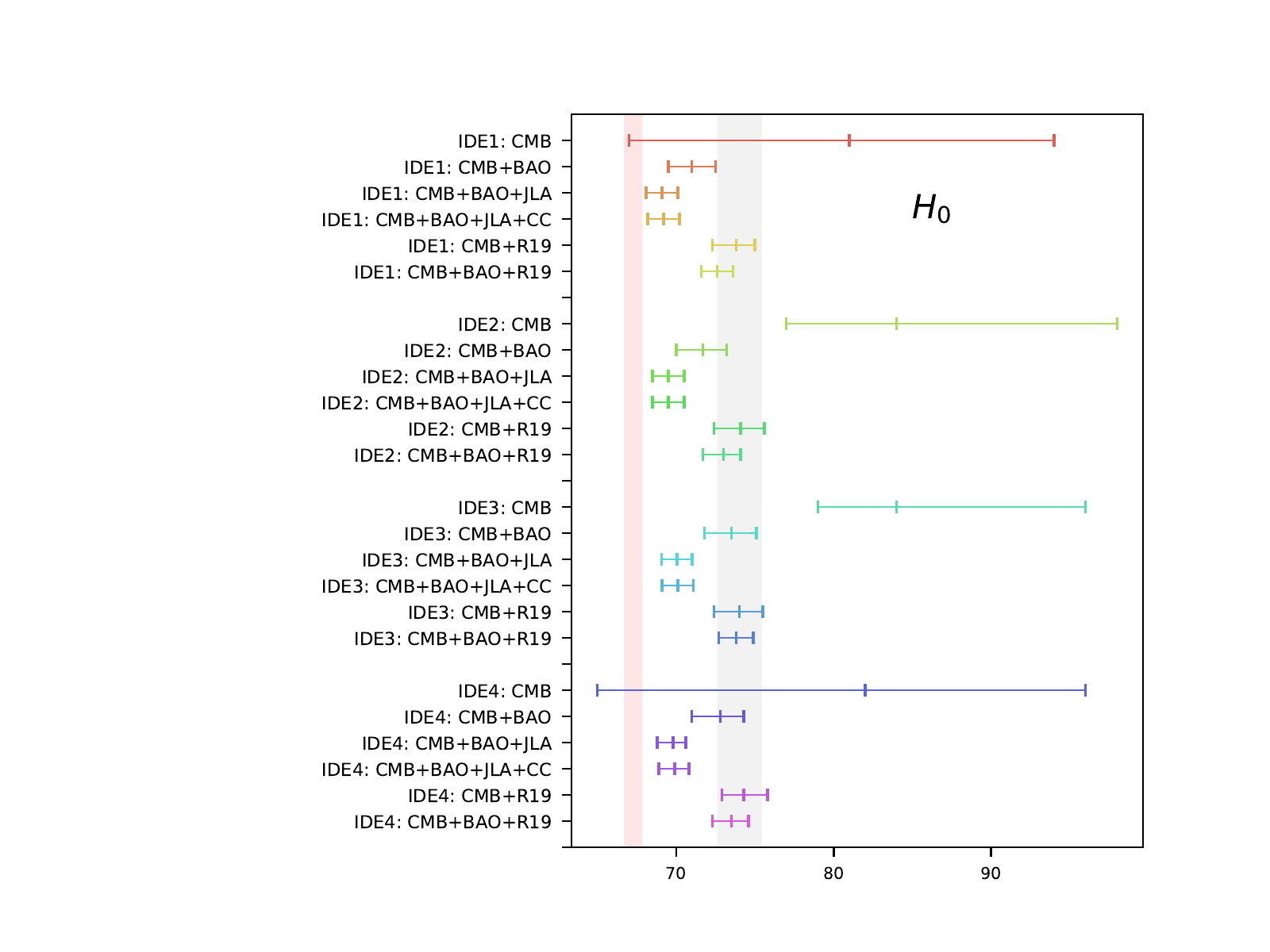}
\caption{
{\it{Whisker plot with the 68\%~CL constraints on the Hubble constant for all interacting 
models and all combination of datasets considered in this work. The grey vertical band 
corresponds to the R19 value for the Hubble constant, $H_0$, as measured by SH0ES 
in~\cite{Riess:2019cxk}, and the red vertical band is the one estimate by the Planck 2018 release~\cite{Aghanim:2018eyx}.}}}
\label{fig-whiskerH0}
\end{center}
\end{figure*}

\subsection{IDE1: Interacting dark energy with $w_x = w_0 a[1-\log(a)]$}
\label{subsec-ide1}

The summary of the observational constraints for this interaction scenario using 
different 
observational datasets is presented in Table \ref{tab:modelI}, while the 2-D contour 
plots 
and the 1-D marginalized posterior distribution are shown in Figs.~\ref{fig-contour-ide1} 
and~\ref{fig-contour-ide1-R19}.  We mention that in the figures we do not include the 
sole CMB   case, since its parameter space is   larger than the other 
datasets, however we note that the qualitative nature of the correlations between the 
parameters for CMB alone and other cases are   similar. 
Moreover, we notice that the addition of CC to the CMB+BAO+JLA combination does not add 
extra constraining power, and hence the constraints from CMB+BAO+JLA and 
CMB+BAO+JLA+CC
are actually the same in the fourth and fifth columns of Table~\ref{tab:modelI}.

From the results we observe that for both CMB and CMB+BAO $\xi =0$ is consistent within 
68\% CL. After the inclusion of JLA and JLA+CC to the combined dataset CMB+BAO, we find 
that an interaction of about $\xi=0.003 \pm 0.002$ is suggested at 68\% CL. In addition, 
if  we combine CMB with R19 (we can safely do it since the tension on $H_0$ is 
less than $2\sigma$ as we can see in Fig.~\ref{fig-whiskerH0}) we find an improved constraint (statistically very mild) but always in agreement with $\xi=0$, while combining CMB+BAO+R19 gives slightly relaxed constraints  Therefore, we 
conclude  that for CMB+BAO+JLA, CMB+BAO+JLA+CC, CMB+R19 and CMB+BAO+R19,  $\xi 
\neq 0$ at 1$\sigma$, however within 95\% CL, $\xi$ is   consistent with zero. 

Concerning the current value of the dark energy equation of state $w_0$, 
for all  combination of datasets it always lies in the phantom regime at more than 
two/three standard deviations. If we compare these results with those without interaction 
obtained in \cite{Yang:2018qmz}, we see that they are perfectly in agreement and very 
robust, even in those cases where an interaction different from zero is favoured.

Regarding the estimated values of $\Omega_{m0}$, one can clearly see that for CMB alone case, $\Omega_{m0}$ is really small compared to what Planck estimates \cite{Ade:2015xua}. This is due to the geometrical degeneracy existing between $w_0$, $\Omega_{m,0}$ and $H_0$. Since a phantom dark energy equation of state is preferred by the Planck data, the position of the peaks in the high multipoles damping tail will be shifted, and we need a lower value for the matter density and an higher Hubble constant to put them back in the measured position. 
When BAO data are added to CMB, the mean value of $\Omega_{m0}$ increases with reduced error bars, because this dataset fixes very well the amount of matter density in the universe. For this reason, all the other dataset combinations we considered have $\Omega_{m0}$ greater than the CMB alone case, even if
always lower than the Planck one~\cite{Ade:2015xua}.

Finally, concerning the estimation of   $H_0$, we see that for CMB data alone it takes 
a very high mean value compared to the $\Lambda$CDM-based Planck's
estimation \cite{Aghanim:2018eyx} and the error bars are quite large (as one can see 
$H_0 =    81_{-   14}^{+   13}$ at 68\% CL for CMB alone). This is an implication of  
the strong anti-correlation   between $w_0$ and $H_0$. However, when the BAO 
data are added to CMB, the error bars on $H_0$ are significantly decreased and its 
estimated mean value shifts towards a lower value ($H_0 = 71.0\pm 1.5$ at 68\% CL for 
CMB+BAO), i.e. perfectly in agreement with the direct 
measurements~\cite{Riess:2016jrr,R18, Riess:2019cxk} within 2$\sigma$. The inclusion of 
JLA (or JLA+CC) to CMB+BAO further decreases the error bars on $H_0$ and further shifts 
its lower value, increasing the $H_0$ tension, but still less than $3\sigma$.

\begingroup                                                     
\squeezetable                                               
\begin{center}                          
\begin{table*}[!]  
\scalebox{0.9}         
   {\begin{tabular}{cccccccccccc}                    
               
\hline\hline

Parameters & CMB & CMB+BAO & CMB+BAO+JLA & CMB+BAO+JLA+CC & CMB+R19 & CMB+BAO+R19 \\ 
\hline
$\Omega_c h^2$ & $    0.1214_{-    0.0022-    0.0040}^{+    0.0017+    0.0042}$ &  $    
0.1197_{-    0.0012-    0.0024}^{+    0.0012+    0.0025}$ & $    0.1191_{-    0.0011-    
0.0023}^{+    0.0011+    0.0022}$ & $    0.1191_{-    0.0012-    0.0024}^{+    0.0012+    
0.0024}$  & $    0.12061377_{-    0.0014-    0.0039}^{+    0.0015+    0.0029}$  & $    
0.1199_{-    0.0013-    0.0023}^{+    0.0012 +    0.0024}$  \\

$\Omega_b h^2$ & $    0.02218_{-    0.00016-    0.00034}^{+    0.00016+    0.00032}$ & $  
 0.02221_{-    0.00015-    0.00026}^{+    0.00014+    0.00029}$ & $    0.02222_{-    
0.00014-    0.00027}^{+    0.00014+    0.00027}$ & $    0.02223_{-    0.00013-    
0.00026}^{+    0.00014+    0.00027}$ & $    0.02215_{-    0.00013-    0.00026}^{+    
0.00014+    0.00027}$ & $    0.02220_{- 0.00016-    0.00028}^{+    0.00014+    0.00029}$ 
\\

$100\theta_{MC}$ & $    1.04034_{-    0.00034-    0.00072}^{+    0.00035+    0.00071}$ & 
$ 
   1.04045_{-    0.00030-    0.00060}^{+    0.00031+    0.00058}$ & $    1.04050_{-    
0.00031-    0.00060}^{+    0.00034+    0.00058}$  & $    1.04051_{-    0.00032-    
0.00065}^{+    0.00032+    0.00066}$  & $    1.04036_{-    0.00034-    0.00073}^{+    
0.00034+    0.00068}$  & $    1.04042_{-    0.00030-    0.00060}^{+    0.00030+    
0.00060}$ \\

$\tau$ & $    0.079_{-    0.018-    0.035}^{+    0.018+    0.034}$ & $    0.088_{-    
0.018-    0.035}^{+    0.018+    0.034}$  & $    0.095_{-    0.017-    0.033}^{+    
0.016+ 
   0.032}$  & $    0.094_{-    0.017-    0.033}^{+    0.018+    0.033}$  & $    0.082_{-  
 0.016-    0.036}^{+    0.017+    0.033}$  & $    0.086_{-    0.017-    0.033}^{+    
0.017+    0.033}$ \\

$n_s$ & $    0.9729_{-    0.0045-    0.0092}^{+    0.0045+    0.0089}$ & $    0.9748_{-   
 0.0043-    0.0084}^{+    0.0042+    0.0084}$ & $    0.9762_{-    0.0042-    0.0081}^{+   
 
0.0042+    0.0082}$ & $    0.9764_{-    0.0045-    0.0076}^{+    0.0042+    0.0079}$ &  $ 
 0.9723_{-    0.0049-    0.0094}^{+    0.0051+    0.0098}$  & $    0.9741_{-    0.0040-  
 0.0084}^{+    0.0040+    0.0084}$ \\

${\rm{ln}}(10^{10} A_s)$ & $    3.103_{-    0.035-    0.070}^{+    0.035+    0.066}$ & $  
 3.119_{-    0.034-    0.068}^{+    0.035+    0.066}$ & $    3.131_{-    0.032-    
0.064}^{+    0.032+    0.066}$  & $    3.129_{-    0.034-    0.067}^{+    0.033+    
0.065}$ & $    3.108_{-    0.031-    0.072}^{+    0.034+    0.063}$ & $ 3.115_{-    
0.033- 
   0.064}^{+    0.035+    0.065}$ \\

$w_0$ & $  <-1.53\,<-1.08$ & $   -1.249_{-    0.055-    0.12}^{+    0.064+    0.12}$ & $  
 -1.166_{-    0.038-    0.074}^{+    0.038+    0.072}$ & $   -1.168_{-    0.034-    
0.081}^{+    0.041+    0.075}$  & $   -1.341_{-    0.047-    0.11}^{+    0.058+    0.09}$ 
 
& $   -1.292_{-    0.042-    0.089}^{+    0.051+    0.083}$ \\

$\xi$ & $    < 0.0071\, <0.015 $ & $    0.0024_{-    0.0016}^{+    0.0015}\, <0.0047$  & 
$ 
 <0.0027\,<0.0038 $  & $    0.0021_{-    0.0013}^{+    0.0014}\,<0.0039$ & $    0.0024_{- 
 0.0021}^{+    0.0009}\,<0.0053$ & $    0.0027_{-    0.0017}^{+    0.0016}\,<0.0051$  \\

$\Omega_{m0}$ & $    0.212_{-    0.071-    0.085}^{+    0.023+    0.132}$ & $    0.277_{- 
 0.011-    0.023}^{+    0.012+    0.022}$ & $    0.295_{-    0.0085-    0.018}^{+    
0.0086+    0.017}$  & $    0.294_{-    0.0086-    0.018}^{+    0.0085+    0.018}$  & $    
0.261_{-    0.011-    0.022}^{+    0.011+    0.022}$  & $    0.2683_{-    0.0081-    
0.016}^{+    0.0076+    0.016}$ \\

$\sigma_8$ & $    0.956_{-    0.044-    0.169}^{+    0.100+    0.123}$ & $    0.859_{-    
0.022-    0.043}^{+    0.022+    0.043}$ & $    0.835_{-    0.017-    0.033}^{+    0.017+ 
 0.032}$  & $    0.835_{-    0.018-    0.036}^{+    0.018+    0.037}$ & $    0.886_{-    
0.023-    0.036}^{+    0.020+    0.037}$ & $    0.872_{-    0.019-    0.038}^{+    0.020+ 
 0.038}$ \\

$H_0$ & $ 84_{-    7-   21}^{+   14+   16} $ &  $   71.7_{-    1.7-    3.1}^{+    1.5+    
3.3}$ & $   69.5_{-    1.0-    1.9}^{+    1.0+    2.0}$  & $   69.5_{-    1.0-    1.9}^{+ 
 1.0+    2.0}$  & $   74.1_{-    1.7-    2.8}^{+    1.5+    2.9}$  & $   73.0_{-    1.3- 
 2.1}^{+    1.1+    2.2}$ \\

$S_8$ & $   0.811_{-   0.048-   0.078}^{+   0.045+   0.069}$ & $   0.817_{-    0.017-    
0.069}^{+    0.020+    0.038}$ & $   0.820_{-    0.018-    0.065}^{+ 0.020 +  0.037}$ & $ 
 0.819_{-    0.019-    0.060}^{+    0.020+    0.036}$  & $   0.826_{-    0.016-    
0.035}^{+    0.017+    0.034}$  & $   0.824_{-    0.015-    0.031}^{+    0.016+    
0.030}$\\

\hline\hline                                                   
\end{tabular} 
}
\caption{68\% and 95\% confidence-level constraints on the interacting scenario IDE2 
with the dark energy equation of state  $w_x(a)=w_0 a \exp(1-a)$ (Model II) for various 
observational datasets. Here $\Omega_{m0}$ is the present value of the 
total matter density parameter $\Omega_m = \Omega_{b}+\Omega_{c}$, and $H_0$ is in   
units of km/sec/Mpc.}\label{tab:modelII}                                   
\end{table*}

\end{center}

\endgroup

\begin{figure*}[!]
\includegraphics[width=0.68\textwidth]{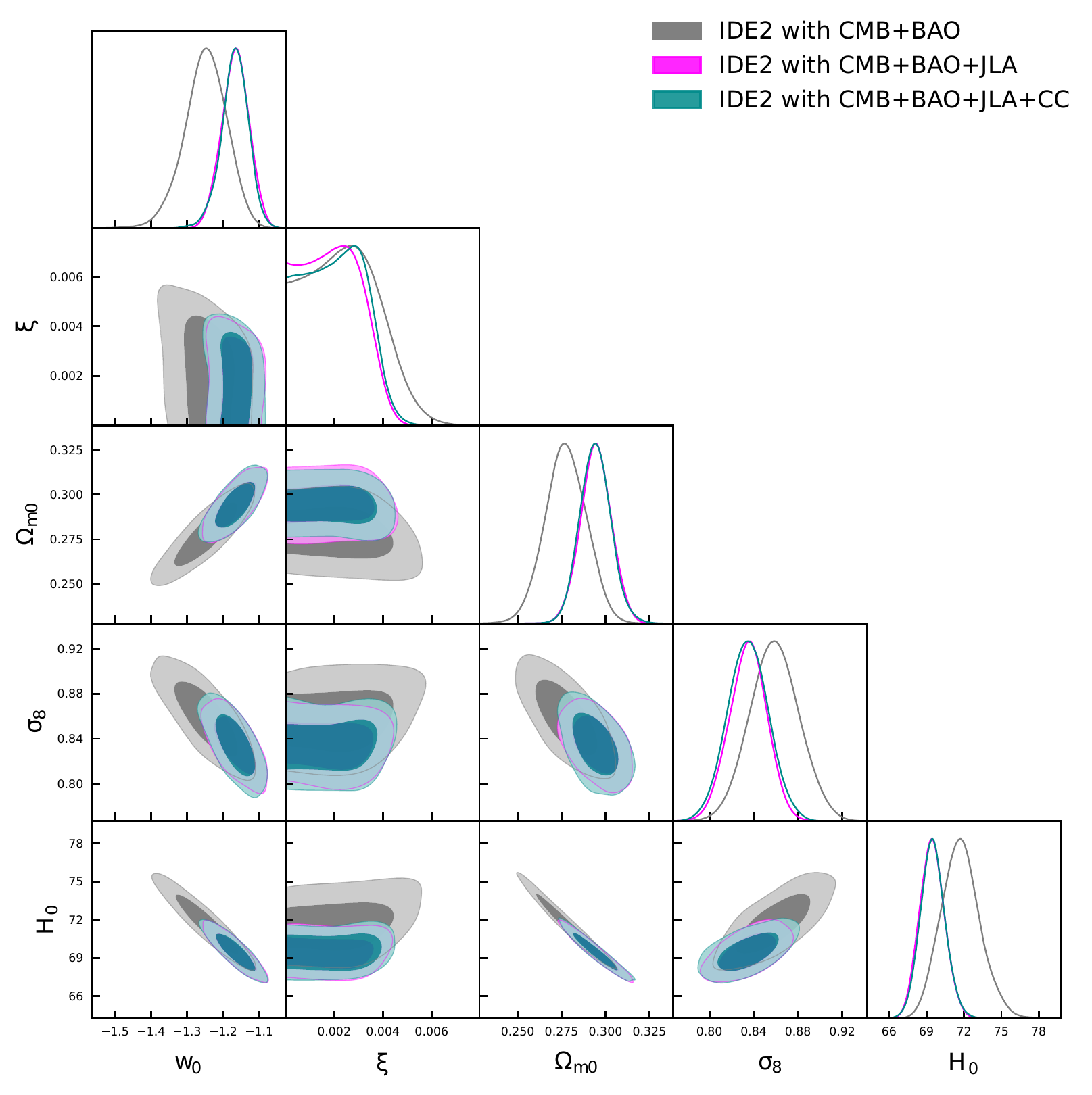}
\caption{{\it{The 68\% and 95\% CL contour plots between various combinations of the 
model parameters of scenario IDE2, using different observational astronomical datasets. 
Additionally we  display the one-dimensional 
marginalized posterior distributions of some free parameters.}}
 }
\label{fig-contour-ide2}
\end{figure*}

\begin{figure*}[!]
\includegraphics[width=0.68\textwidth]{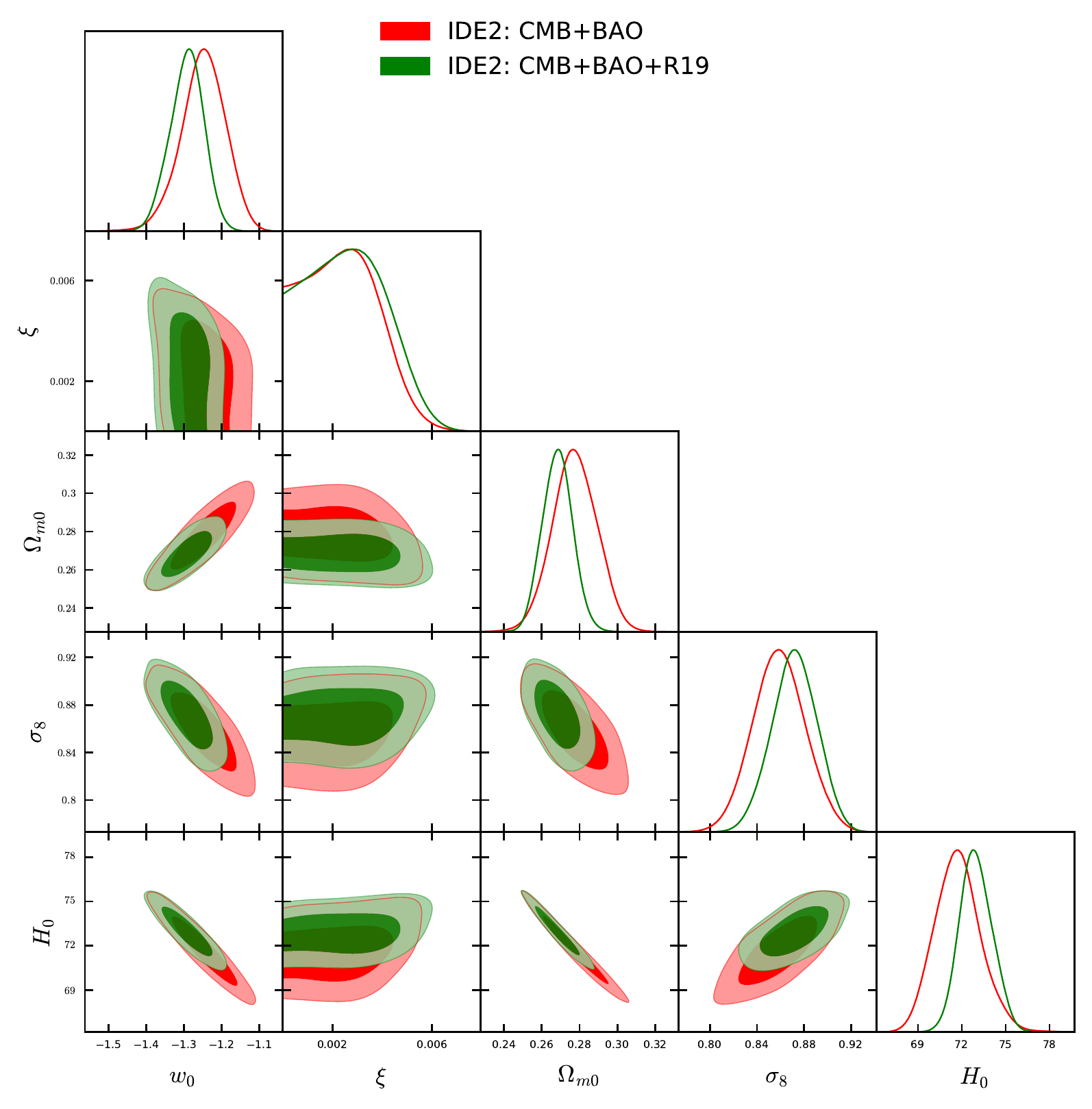}
\caption{{\it{The 68\% and 95\% CL contour plots between various combinations of the model 
parameters of scenario IDE2 using only   the CMB+BAO and CMB+BAO+R19 datasets, and the 
corresponding
one-dimensional marginalized posterior distributions.}} }
\label{fig-contour-ide2-R19}
\end{figure*}

\subsection{IDE2: Interacting dark energy with $w_x (a) = w_0 a \exp(1-a)$}

The observational summary for this interaction scenario is displayed in Table 
\ref{tab:modelII}, while the 2-D contour plots and 1-D parameter distributions are shown 
in Figs.~\ref{fig-contour-ide2} and~\ref{fig-contour-ide2-R19}. From the analyses we 
deduce that for CMB data alone  the non-interacting case $\xi = 0$ is consistent within 
68\% CL, however  when BAO data are added to CMB  then an indication of interaction is 
found at more than 68\% CL. Surprisingly when JLA data are added to the previous dataset 
CMB+BAO, then we again find that $\xi =0$ consistent within 68\% CL. Moreover, for the 
remaining datasets the indication of an interaction is still present at more than 
1$\sigma$. We note that due to the addition of R19 with CMB and CMB+BAO, slight improvements in the coupling parameter $\xi$ are observed, although such improvements are statistically very mild. In fact, R19 sets a lower bound on $\xi$ for both CMB+R19 and 
CMB+BAO+R19. The reason is that the inclusion of R19 with CMB and CMB+BAO breaks the degeneracies between $H_0$ and other cosmological parameters.

Concerning the dark energy equation-of-state parameter at present, for all the datasets a 
phantom value    $w_0< -1$ is always 
supported for more than 95\% CL. 
Hence, in summary, as we can from the results,   most of 
the datasets indicate a non-zero interaction  together with the   
existence of a phantom dark energy. A similar observation on the estimated values of $\Omega_{m0}$, specially for its lower value for CMB alone analysis, is found as in IDE1. 

Regarding the estimations of the Hubble parameter $H_0$, we see that for CMB alone $H_0$ 
acquires a very high value with very large error bars compared to the Planck one within 
minimal $\Lambda$CDM model \cite{Aghanim:2018eyx} (in particular  $H_0 = 
84_{-7}^{+   14}$ at 68\% CL with  CMB alone). This is due to the strong correlation 
between $w_0$ and $H_0$.  
When external datasets are added, as for instance BAO, JLA, CC, R19, and their 
combinations, the estimations of $H_0$ decrease with significant reduction in the error 
bars, but they can still relieve the tension with \cite{Riess:2019cxk} within $3$  standard deviations.

\subsection{IDE3: Interacting dark energy with $w_x(a)=w_0a[1+\sin(1-a)]$}
\label{subsec-ide1}

\begin{figure*}[!]
\includegraphics[width=0.68\textwidth]{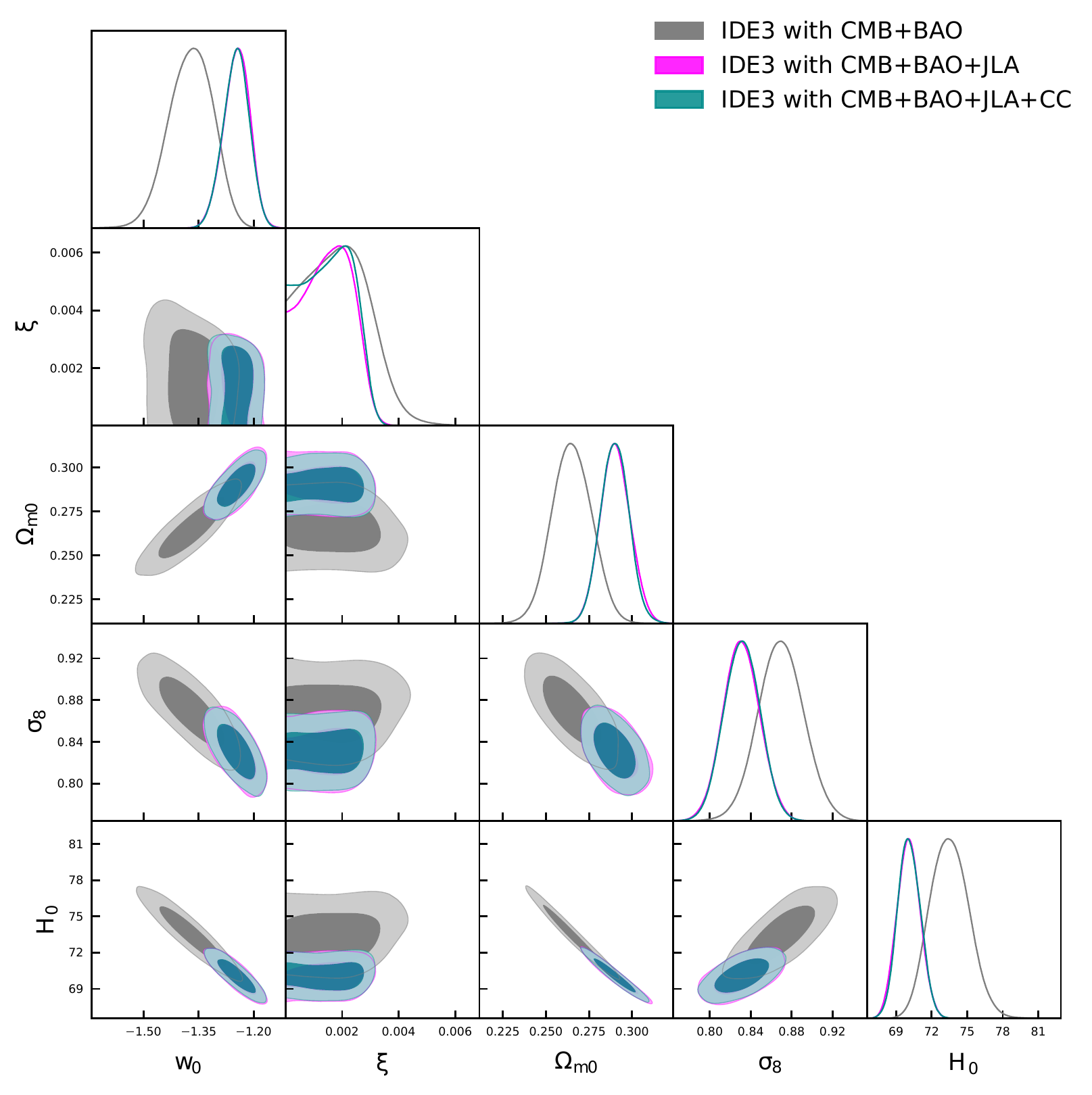}
\caption{{\it{The 68\% and 95\% CL contour plots between various combinations of the 
model parameters of scenario IDE3, using different observational astronomical datasets. 
Additionally we  display the one-dimensional 
marginalized posterior distributions of some free parameters.}}
 }
\label{fig-contour-ide3}
\end{figure*}

\begin{figure*}[!]
\includegraphics[width=0.68\textwidth]{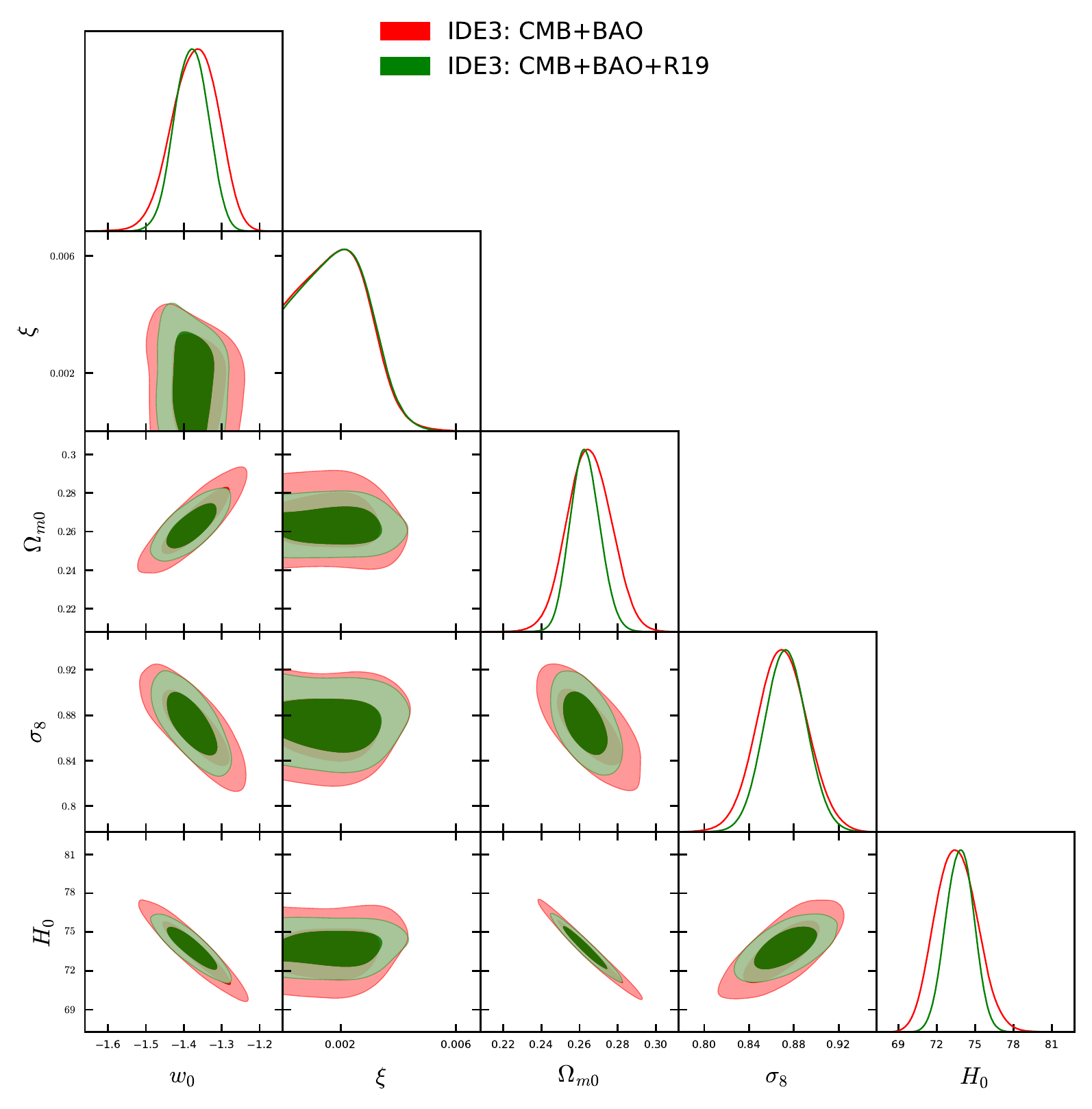}
\caption{{\it{The 68\% and 95\% CL contour plots between various combinations of the 
model parameters of scenario IDE3 using only   the CMB+BAO 
and CMB+BAO+R19 datasets, and the corresponding
one-dimensional 
marginalized posterior distributions.}}   }
\label{fig-contour-ide3-R19}
\end{figure*}

The summary of the observational constraints for this interaction scenario using 
different 
observational datasets is presented in Table \ref{tab:modelIII} and in 
Figs.~\ref{fig-contour-ide3} and \ref{fig-contour-ide3-R19} we show the 2-D contour 
plots and 1-D posterior distributions for some of the free parameters and dataset 
combinations. 

Concerning the coupling parameter our analysis reveals  some interesting 
features. In particular, as we can 
see, for CMB data alone we have $\xi =0$  within 68\% CL and hence it is 
consistent 
with a non-interacting cosmology. Nevertheless, as soon as external datasets, namely 
BAO, JLA, CC or R19 are added in different combinations (such as CMB+BAO, CMB+BAO+JLA, 
CMB+BAO+JLA+CC and CMB+R19) we see that  a non-zero
interaction   is  favoured  at more than 1$\sigma$. However, we mention that  within 
95\% 
CL these combinations of observational datasets  allow for a non-interacting cosmology. We note that the R19 has similar effects when combined with CMB and CMB+BAO, as already reported previously for IDE2.

\begingroup                              
\squeezetable                           
\begin{center}                       
\begin{table*}     [!]    
\scalebox{0.9}        
 { \begin{tabular}{cccccccccccccccccc}                        
\hline\hline

Parameters & CMB & CMB+BAO & CMB+BAO+JLA & CMB+BAO+JLA+CC & CMB+R19 & CMB+BAO+R19 \\ 
\hline

$\Omega_c h^2$ & $    0.1211_{-    0.0015-    0.0031}^{+    0.0015+    0.0031}$ & $    
0.1202_{-    0.0013-    0.0023}^{+    0.0011+    0.0025}$ & $    0.1196_{-    0.0012-    
0.0023}^{+    0.0011+    0.0022}$ & $    0.1196_{-    0.0012-    0.0021}^{+    0.0011+    
0.0022}$  & $    0.1212_{-    0.0016-    0.0029}^{+    0.0014+    0.0030}$  & $    
0.1203_{-    0.0013-    0.0023}^{+  0.0012 +    0.0024}$ \\

$\Omega_b h^2$ & $    0.02216_{-    0.00016-    0.00037}^{+    0.00019+    0.00035}$ & $  
 
 0.02216_{-    0.00014-    0.00027}^{+    0.00014+    0.00027}$ & $    0.02218_{-    
0.00016-    0.00028}^{+    0.00015+    0.00028}$ & $    0.02218_{-    0.00013-    
0.00027}^{+    0.00013+    0.00026}$ & $    0.02208_{-    0.00016-    0.00030}^{+    
0.00016+    0.00029}$ & $    0.02216_{- 0.00014- 0.00027}^{+    0.00014+    0.00027}$ \\

$100\theta_{MC}$ & $    1.04033_{-    0.00035-    0.00070}^{+    0.00039+    0.00066}$ & 
$ 
   1.04035_{-    0.00030-    0.00062}^{+    0.00033+    0.00060}$ & $    1.04043_{-    
0.00030-    0.00059}^{+    0.00030+    0.00058}$ &  $    1.04041_{-    0.00029-    
0.00056}^{+    0.00030+    0.00058}$ & $    1.04019_{-    0.00036-    0.00070}^{+    
0.00039+    0.00066}$ & $    1.04036_{-    0.00030-    0.00063}^{+    0.00033+    
0.00060}$ \\

$\tau$ & $    0.079_{-    0.018-    0.035}^{+    0.017+    0.035}$ & $    0.089_{-    
0.017-    0.035}^{+    0.019+    0.034}$ & $    0.098_{-    0.018-    0.035}^{+    0.018+ 
 
  0.036}$ & $    0.098_{-    0.017-    0.033}^{+    0.017+    0.034}$  & $    0.083_{-    
0.020-    0.037}^{+    0.020+    0.037}$ & $    0.089_{-    0.017-    0.034}^{+    0.019+ 
 
  0.033}$ \\

$n_s$ & $    0.9725_{-    0.0045-    0.0086}^{+    0.0046+    0.0087}$ & $    0.9740_{-   
 
0.0040-    0.0080}^{+    0.0040+    0.0076}$ & $    0.9761_{-    0.0042-    0.0080}^{+    
0.0042+    0.0080}$ & $    0.9764_{-    0.0040-    0.0077}^{+    0.0039+    0.0077}$  & $ 
 
  0.9713_{-    0.0051-    0.0082}^{+    0.0044+    0.0090}$ & $    0.9739_{-    0.0040-   
 
0.0081}^{+    0.0040+    0.0076}$ \\

${\rm{ln}}(10^{10} A_s)$ & $    3.102_{-    0.035-    0.065}^{+    0.033 +    0.068}$ & $ 
 
  3.120_{-    0.034-    0.068}^{+    0.037+    0.064}$ &  $    3.135_{-    0.035-    
0.069}^{+    0.034+    0.069}$ & $    3.136_{-    0.033-    0.064}^{+    0.032+    
0.065}$ 
& $    3.111_{-    0.038-    0.068}^{+    0.040+    0.067}$ & $    3.120_{-    0.033-    
0.064}^{+    0.033+    0.063}$ \\

$w_0$ & $ <-1.61\,<-1.17$ & $   -1.371_{-    0.058-    0.12}^{+    0.066+    0.12}$ & $   
-1.246_{-    0.033-    0.072}^{+    0.040+    0.065}$ & $   -1.248_{-    0.035-    
0.070}^{+    0.037+    0.065}$ & $   -1.406_{-    0.053-    0.091}^{+    0.051+    
0.093}$ 
& $   -1.380_{-    0.044-    0.084}^{+    0.044+    0.083}$ \\

$\xi$ & $ <0.0045\,<   0.0080$ & $    0.0019_{-    0.0012}^{+    0.0011}\,<0.0036$ & $    
0.0015_{-    0.0008}^{+    0.0010}\,<0.0027$ & $    0.0015_{-    0.0008}^{+    
0.0011}\,<0.0027$  & $    0.0023_{-    0.0011}^{+    0.0014}\,<0.0039$ & $    0.0019_{-   
0.0012}^{+    0.0011}\,<0.0036$ \\

$\Omega_{m0}$ & $    0.210_{-    0.065-    0.079}^{+    0.020+    0.12}$ & $    0.265_{-  
 0.011-    0.022}^{+    0.012+    0.022}$ & $    0.290_{-    0.009-    0.016}^{+    
0.009+ 
   0.017}$ & $    0.290_{-    0.008-    0.016}^{+    0.008+    0.016}$  & $    0.263_{-   
 
0.011-    0.020}^{+    0.011+    0.021}$ & $    0.2629_{- 0.0078-    0.015}^{+    0.0078+ 
 
  0.016}$ \\

$\sigma_8$ & $    0.956_{-    0.039-    0.164}^{+    0.096+    0.118}$ & $    0.870_{-    
0.023-    0.043}^{+    0.022+    0.044}$ & $    0.831_{-    0.018-    0.034}^{+    0.017+ 
  0.035}$ & $    0.831_{-    0.018-    0.034}^{+    0.018+    0.034}$ & $    0.880_{-    
0.020-    0.040}^{+    0.019+    0.040}$ & $    0.873_{-    0.018-    0.036}^{+    0.018+ 
 
  0.037}$ \\

$H_0$ & $ 84_{-  5-   19}^{+   12+   15}$ & $   73.5_{-    1.7-    3.2}^{+    1.6+    
3.2}$ & $   70.06_{-    0.98-    1.8}^{+    0.95+    2.0}$ & $   70.1_{-    1.0-    
1.8}^{+    1.0+    1.9}$ & $   74.0_{-  1.6 -2.8}^{+ 1.5+    2.7}$ & $ 73.8_{-    1.1-    
2.1}^{+    1.1+    2.1}$ \\

$S_8$ & $   0.807_{-   0.044-   0.074}^{+   0.045+   0.069}$ & $   0.803_{-    0.021-    
0.097}^{+    0.025+    0.043}$ & $   0.803_{-    0.025-    0.090}^{+ 0.026 +  0.043}$ & $ 
 
 0.800_{-    0.030-    0.090}^{+    0.028+    0.044}$  & $   0.823_{-    0.016-    
0.032}^{+    0.017+    0.030}$  & $   0.817_{-    0.015-    0.029}^{+    0.015+    
0.030}$\\

\hline\hline                                                    \end{tabular}
}
\caption{68\% and 95\% confidence-level constraints on the interacting scenario IDE3 
with the dark energy equation of state  $w_x(a)=w_0 a[1+\sin(1-a)]$ (Model III) for 
various observational datasets. Here $\Omega_{m0}$ is the present value 
of the total matter density parameter $\Omega_m = \Omega_{b}+\Omega_{c}$, and $H_0$ is in 
  units of km/sec/Mpc.}\label{tab:modelIII}

\end{table*}                        
\end{center}                        
\endgroup

Concerning the current value of the dark energy equation of state $w_0$, we find a 
similar character to what we already found in IDE1 and IDE2. In particular, the results 
show that irrespectively of the observational datasets that we have used in this 
work, $w_0$ remains less than $-1$ at more than 95\% CL, i.e. in the phantom 
region, for the CMB only case, and several standard deviations (more than five) for the 
combinations with the other cosmological probes.
If we compare this Table with the results shown in \cite{Yang:2018qmz} for the same model 
without interaction, we can see that the constraints are very robust, and only the upper 
limit of $w_0$ for the CMB alone case is slightly removed to less phantom values. 
Furthermore we mention that  in this scenario the CC dataset does not improve the 
constraints at all. Let us note that similar to IDE1 and IDE2, for CMB alone analysis, $\Omega_{m0}$ acquires a lower value compared to Planck \cite{Ade:2015xua}. 

Finally, regarding the $H_0$ parameter we again find a similar behaviour to what we 
observed for IDE1 and IDE2. Since to a phantom dark energy equation of state corresponds 
a higher value of the Hubble parameter, due to their negative correlation, the highly 
negative $w_0$ values that we obtain are accompanied with a high value of $H_0$ with 
large asymmetric error bars. Specifically, we find $H_0 = 84_{-5}^{+   12}$  at 68\% CL, 
which is much higher than the recent $\Lambda$CDM-based estimation by Planck 
\cite{Aghanim:2018eyx}, but in agreement with the direct measurements $H_0=73.24\pm1.74$ 
of~\cite{Riess:2016jrr}, $H_0=73.48\pm1.66$ of~\cite{R18} or $H_0=74.03\pm1.42$ 
of~\cite{Riess:2019cxk}. However, after the inclusion of the external datasets such as 
BAO, JLA, CC and R19 we find that $H_0$ decreases with respect to its estimation from CMB 
alone, and additionally its error bars are significantly reduced. 

In summary, we find that the 
alleviation of the $H_0$ tension is more robust in this scenario compared to IDE1 and 
IDE2 
(see also Fig~\ref{fig-whiskerH0}). Indeed, one can notice the estimated values of $H_0$ 
from different combination of observational datasets as follows:
\begin{itemize}
    \item CMB+BAO: $H_0 = 73.5_{-    1.7}^{+    1.6}$ at 68\% CL ($H_0 = 73.5\pm 3.2$ at 
95\% CL);  
    \item CMB+BAO+JLA: $H_0 =   70.06_{- 0.98}^{+ 0.95}$ at 68\% CL ($H_0 = 
70.1_{-1.8}^{+2.0}$, at 95\% CL);
    \item CMB+BAO+JLA+CC: $ H_0=  70.1\pm 1.0$ at 68\% CL ($ H_0=  70.1_{-1.8}^{+1.9}$, 
at 
95\% CL),
\end{itemize}
where the first one is perfectly in agreement with \cite{Riess:2019cxk}, and the last two 
alleviate the tension at about $2\sigma$.

\begingroup                         
\squeezetable                                                                             
 
 \begin{center}                        
\begin{table*} [!]
\scalebox{0.9}
{\begin{tabular}{cccccccccccccccccc}

\hline\hline

Parameters & CMB & CMB+BAO & CMB+BAO+JLA & CMB+BAO+JLA+CC & CMB+R19 & CMB+BAO+R19 \\ 
\hline

$\Omega_c h^2$ & $ 0.1212_{-    0.0019-    0.0032}^{+    0.0017+    0.0034}$ & $ 
0.1200_{- 
   0.0012-    0.0026}^{+    0.0012+    0.0026}$ & $    0.1192_{-    0.0011-    0.0022}^{+ 
 
  0.0011+    0.0022}$ & $    0.1193_{-    0.0011-    0.0020}^{+    0.0010+    0.0022}$ & 
$ 
   0.1208_{-    0.0013-    0.0024}^{+    0.0015+    0.0022}$ & $    0.1201_{-    0.0012-  
 
 0.0026}^{+    0.0012+    0.0026}$ \\

$\Omega_b h^2$ & $    0.02214_{-    0.00018-    0.00034}^{+    0.00018+    0.00034}$ & $  
 
 0.02217_{- 0.00015-    0.00027}^{+    0.00014+    0.00027}$ & $    0.02222_{-    
0.00014- 
   0.00028}^{+    0.00013 +    0.00029}$ & $    0.02222_{-    0.00014-    0.00027}^{+    
0.00015+    0.00027}$ & $    0.02212_{-    0.00017-    0.00025}^{+    0.00013+    
0.00028}$  & $    0.02217_{-    0.00015-    0.00028}^{+    0.00014+    0.00027}$  \\

$100\theta_{MC}$ & $    1.04027_{-    0.00034-    0.00080}^{+    0.00042+    0.00076}$ & 
$ 
   1.04040_{-    0.00032-    0.00059}^{+    0.00032+    0.00062}$ &  $    1.04046_{-    
0.00026-    0.00057}^{+    0.00032+    0.00056}$ & $    1.04050_{-    0.00030-    
0.00060}^{+    0.00031+    0.00059}$ & $    1.04027_{-    0.00031-    0.00060}^{+    
0.00036+    0.00057}$  & $    1.04040_{-    0.00031-    0.00058}^{+    0.00033+    
0.00060}$ \\

$\tau$ & $    0.079_{-    0.017-    0.034}^{+    0.019+    0.032}$ & $    0.089_{-    
0.017-    0.034}^{+    0.018+    0.033}$ & $    0.096_{-    0.016-    0.034}^{+    0.016+ 
 
  0.034}$ & $    0.096_{-    0.016-    0.036}^{+    0.019+    0.036}$  & $    0.077_{-    
0.018-    0.038}^{+    0.025+    0.035}$ & $    0.088_{-    0.017-    0.033}^{+    0.018+ 
 
  0.033}$  \\

$n_s$ & $    0.9721_{-    0.0047-    0.0087}^{+    0.0044+    0.0092}$ & $    0.9741_{-   
 
0.0042-    0.0082}^{+    0.0042+    0.0081}$ & $    0.9765_{-    0.0044-    0.0080}^{+    
0.0040+    0.0082}$ & $    0.9764_{-    0.0038-    0.0078}^{+    0.0038+    0.0075}$ & $  
 
 0.9721_{-    0.0036-    0.0084}^{+    0.0046+    0.0078}$  & $    0.9739_{-    0.0042-   
 
0.0080}^{+    0.0042+    0.0081}$  \\

${\rm{ln}}(10^{10} A_s)$ & $    3.103_{-    0.031-    0.067}^{+    0.037+    0.063}$ & $  
 
 3.120_{-    0.033-    0.067}^{+    0.035+    0.066}$ & $    3.132_{-    0.032-    
0.064}^{+    0.034+    0.064}$ & $    3.132_{-    0.032-    0.071}^{+    0.037+    
0.069}$ 
& $    3.099_{-    0.034-    0.076}^{+    0.048+    0.067}$ & $    3.118_{-    0.032-    
0.066}^{+    0.035+    0.064}$  \\

$w_0$ & $ <-1.48\,<-1.02$ & $   -1.331_{-    0.056-    0.13}^{+    0.067+    0.12}$ & $   
-1.218_{-    0.029-    0.074}^{+    0.040+    0.067}$ & $   -1.222_{-    0.033-    
0.072}^{+    0.038+    0.065}$ & $   -1.394_{-    0.048-    0.092}^{+    0.045+    
0.098}$ 
 & $   -1.353_{-    0.043-    0.091}^{+    0.048+    0.084}$  \\

$\xi$ & $  <0.0041\,<0.0093$ & $  <0.0028\,<0.0042$ & $    0.0017_{-    0.0011}^{+    
0.0011}\,<0.0032$ & $    0.0018_{-    0.0011}^{+    0.0012}\,<0.0033$  & $    <0.0030\,<0.0044$ & $    <0.0030\,<0.0042$ \\

$\Omega_{m0}$ & $    0.229_{-    0.089-    0.096}^{+    0.098+    0.15}$ & $    0.270_{-  
 
 0.012-    0.022}^{+    0.011+    0.023}$ & $    0.2919_{-    0.0080-    0.016}^{+    
0.0078+    0.016}$ & $    0.291_{-    0.0080-    0.016}^{+    0.0079+    0.016}$ & $    
0.260_{-    0.011-    0.023}^{+    0.011+    0.024}$  & $    0.2648_{-    0.0080-    
0.015}^{+    0.0078+    0.016}$ \\

$\sigma_8$ & $    0.93_{-    0.14-    0.18}^{+    0.12+    0.14}$ & $    0.867_{-    
0.023-    0.045}^{+    0.023+    0.046}$ & $    0.832_{-    0.018-    0.032}^{+    0.016+ 
 
  0.034}$ & $    0.834_{-    0.017-    0.033}^{+    0.017+    0.034}$  & $    0.881_{-    
0.020-    0.043}^{+    0.024+    0.039}$ & $    0.874_{-    0.019-    0.037}^{+    0.019+ 
 
  0.038}$  \\

$H_0$ & $   82_{-   17-   21}^{+   14+   16}$ & $   72.8_{-    1.8-    3.0}^{+    1.5+    
3.3}$ &  $   69.8_{-    1.0-    1.8}^{+    0.8+    1.9}$ & $   69.9_{-    1.0-    1.8}^{+ 
 
  0.9+    2.0}$ & $   74.3_{-    1.4-    3.2}^{+    1.5+    3.1}$  & $   73.5_{-    1.2-  
 
 2.1}^{+    1.1+    2.2}$ \\

$S_8$ & $   0.818_{-   0.045-   0.074}^{+   0.041+   0.065}$ & $   0.812_{-    0.019-    
0.066}^{+    0.022+    0.039}$ & $   0.808_{-    0.022-    0.073}^{+ 0.022 +  0.040}$ & $ 
 
 0.810_{-    0.021-    0.070}^{+    0.022+    0.038}$  & $   0.820_{-    0.018-    
0.040}^{+    0.017+    0.036}$  & $   0.821_{-    0.016-    0.033}^{+    0.016+    
0.032}$\\

\hline\hline

\end{tabular}      }

\caption{68\% and 95\% confidence-level constraints on the interacting scenario IDE4 
with the dark energy equation of state  $w_x(a)=w_0a[1+\arcsin(1-a)]$ (Model IV) for 
various observational datasets. Here $\Omega_{m0}$ is the present value 
of the total matter density parameter $\Omega_m = \Omega_{b}+\Omega_{c}$, and $H_0$ is in 
  units of km/sec/Mpc.}\label{tab:modelIV}

\end{table*}                          
\end{center}                         
\endgroup

\begin{figure*}[!]
\includegraphics[width=0.70\textwidth]{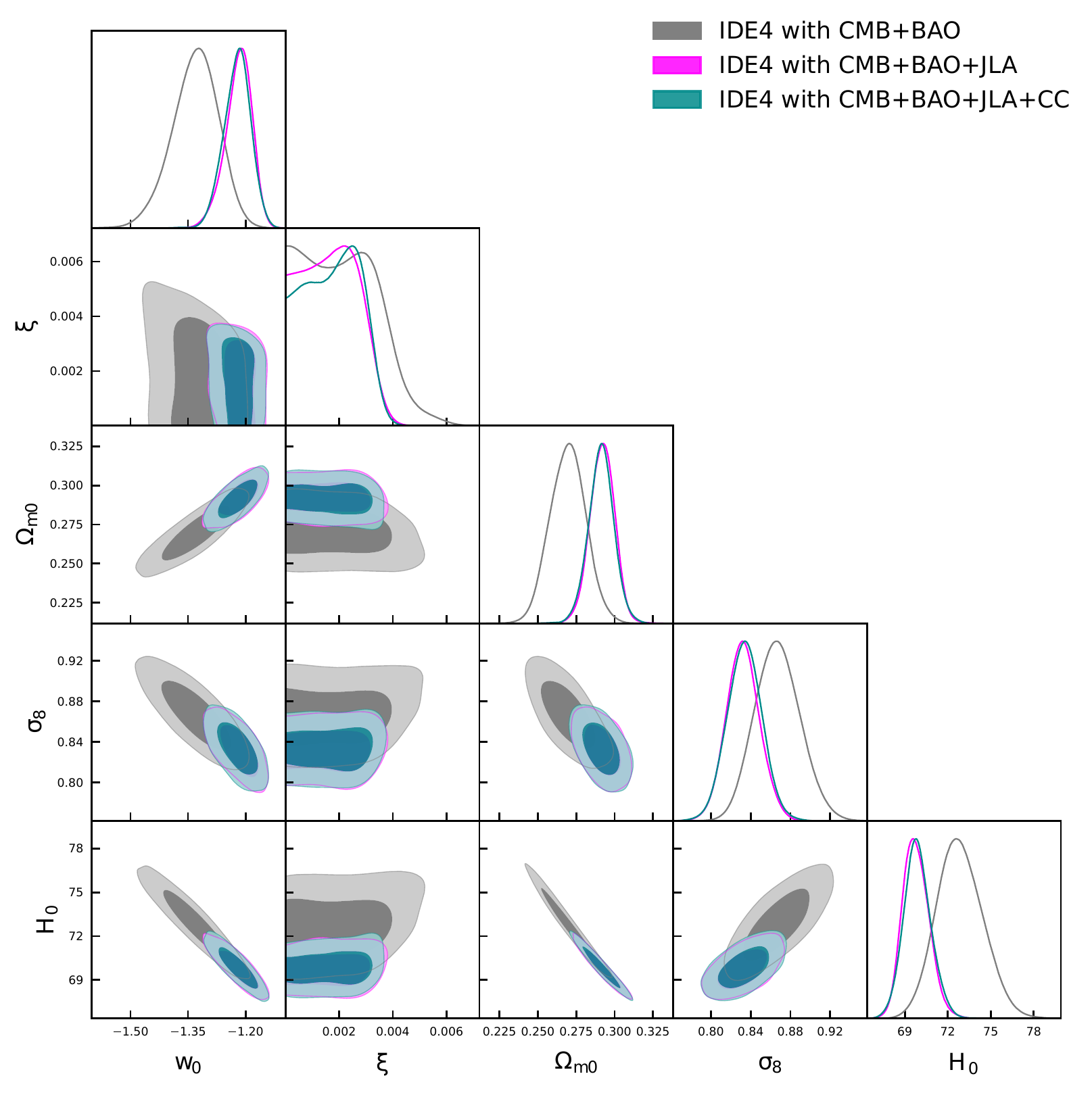}
\caption{{\it{The 68\% and 95\% CL contour plots between various combinations of the 
model parameters of scenario IDE4, using different observational astronomical datasets. 
Additionally we  display the one-dimensional 
marginalized posterior distributions of some free parameters.}} }
\label{fig-contour-ide4}
\end{figure*}    

\begin{figure*}[!]
\includegraphics[width=0.68\textwidth]{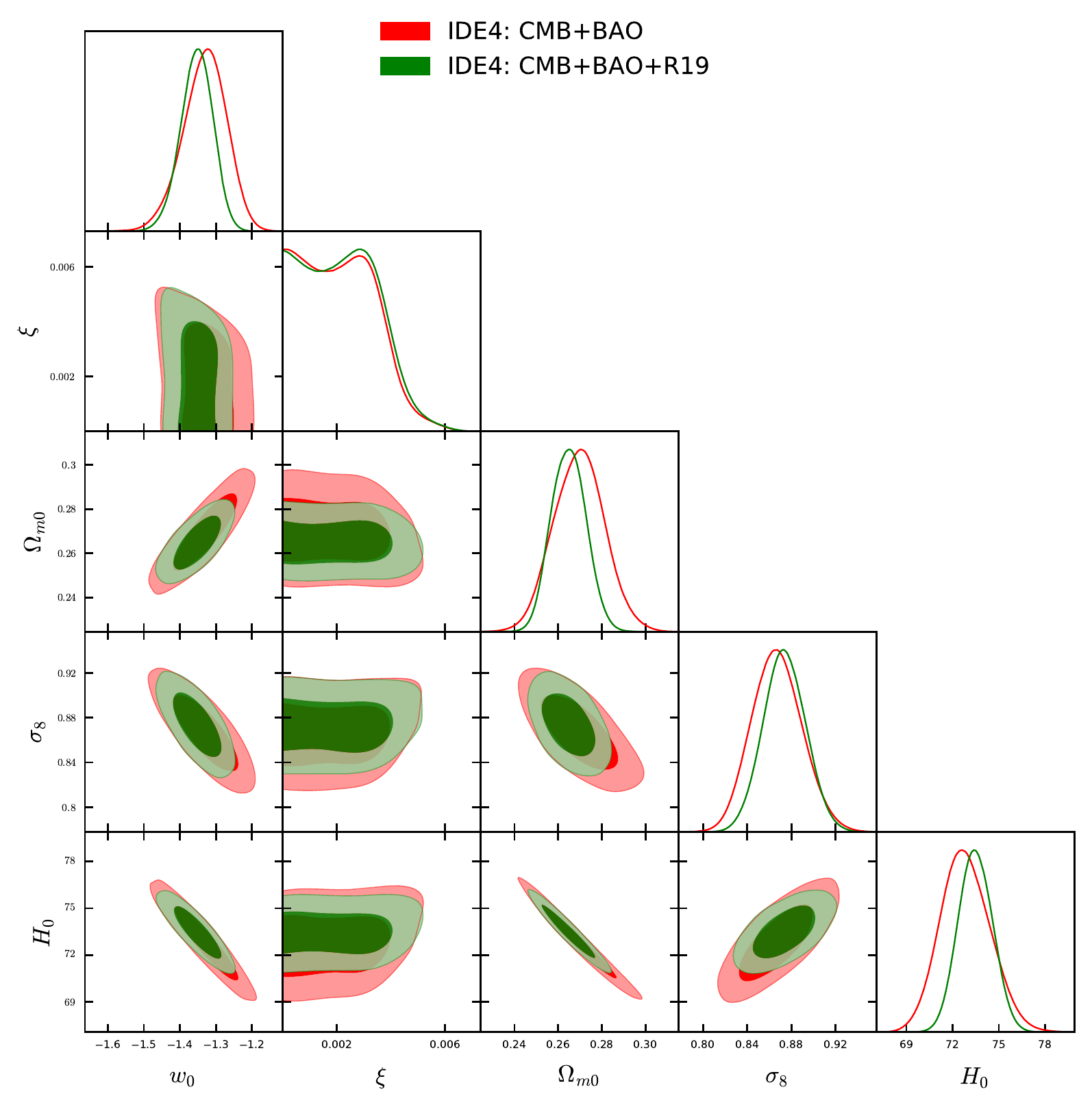}
\caption{{\it{The 68\% and 95\% CL contour plots between various combinations of the 
model parameters of scenario IDE4 using only   the CMB+BAO 
and CMB+BAO+R19 datasets, and the corresponding
one-dimensional 
marginalized posterior distributions.}}   }
\label{fig-contour-ide4-R19}
\end{figure*}

\subsection{IDE4: Interacting dark energy with $w_x(a)=w_0a[1+\arcsin(1-a)]$}
\label{subsec-ide1}

The summary of the observational constraints for this   interacting scenario 
using different 
observational datasets is displayed in Table~\ref{tab:modelIV} and in 
Figs.~\ref{fig-contour-ide4} and \ref{fig-contour-ide4-R19} we present the 2-D contour 
plots and the 1-D posterior distributions.
The behaviour of this interaction scenario has some similarities to that of IDE1. Looking 
at the results we find also in this case that for the analysis with CMB only and CMB+BAO 
datasets, the coupling parameter is consistent with $\xi =0$ within the 68\% CL. The 
addition of JLA or CC, namely the combinations of datasets CMB+BAO+JLA and 
CMB+BAO+JLA+CC, gives instead an indication for $\xi \neq 0$ at more than 68\% 
CL, but always in agreement with zero within 2$\sigma$. We mention that similarly to the previous IDE1 scenarios, here we also see that the inclusion of R19 with CMB and CMB+BAO improves the constraints on $\xi$, however it gives just an upper bound on it. 

Concerning the dark energy equation-of-state parameter at present we extract 
similar 
conclusion to the previous interacting scenarios IDE3, namely here too we find that 
$w_0 < -1$ at more than 95\% CL for the CMB only case, and several standard deviations 
for 
its combination with the external datasets. Furthermore, for this scenario the CMB only 
case has a slightly less phantom
$w_0$  than the same case without interaction, as can be seen in 
\cite{Yang:2018qmz}. As already reported in earlier IDE models, $\Omega_{m0}$ for CMB alone case obtains a lower value, contrary to Planck observations \cite{Ade:2015xua}. 

Now we focus on the trend on  the Hubble parameter $H_0$. For the datasets we use, in 
this 
case it is again  anti-correlated with $w_0$, as we can see in 
Fig.~\ref{fig-contour-ide4}. We note that similar to the previous interaction scenarios, 
the CMB only fit returns very high value of $H_0$ with large error bars, that are reduced 
after 
the inclusion of the external datasets such as BAO, JLA and CC. For this scenario we also
conclude that the tension with the direct measurements \cite{Riess:2016jrr, R18, 
Riess:2019cxk} is 
solved for CMB and CMB+BAO cases, while with the addition of JLA and JLA+CC it is at 
about 
$2\sigma$. For this reason we can safely add the R19 measurement to the CMB and CMB+BAO, 
and we show the results in the last two columns of Table~\ref{tab:modelIV}.

\section{Bayesian evidence}
\label{sec-bayesian}

In this section we compute the Bayesian evidences of all the examined  interacting 
models in order to 
compare their observational 
soundness with respect to some reference model, and in particular with   $\Lambda$CDM 
cosmology.  We use the 
publicly available 
code \texttt{MCEvidence} \cite{Heavens:2017hkr,Heavens:2017afc} 
     \begin{table} [ht]             
\begin{tabular}{ccc}                
\hline\hline

$\ln B_{ij}$ & ~~~~~~~Strength of evidence for model ${M}_i$ \\ \hline
$0 \leq \ln B_{ij} < 1$ & Weak \\
$1 \leq \ln B_{ij} < 3$ & Definite/Positive \\
$3 \leq \ln B_{ij} < 5$ & Strong \\
$\ln B_{ij} \geq 5$ & Very strong \\
\hline\hline                        
\end{tabular}

\caption{Revised Jeffreys scale \cite{Kass:1995loi} that quantifies the comparison of the 
models.} \label{tab:jeffreys}       
\end{table}                                             
\begin{table*} [!]

\begin{center}                    
\begin{tabular}{ccccccccc}                                      \hline\hline              
 
Dataset & Model & $\ln B_{ij}$ & ~~Strength of evidence for reference model $\Lambda$CDM 
\\ 
\hline
CMB & IDE1 & $-2.8$ & Definite/Positive\\
CMB+BAO & IDE1 & $-4.6$ & Strong\\
CMB+BAO+JLA & IDE1 & $-7.3$ & Very Strong\\
CMB+BAO+JLA+CC & IDE1 & $-6.7$ & Very Strong\\
CMB+R19 & IDE1 & $+1.0$ & Weak for IDE1\\
CMB+BAO+R19 & IDE1 & $-0.7$ & Weak\\
\hline
CMB & IDE2 &  $-4.3$ & Strong\\
CMB+BAO & IDE2 & $-4.9$ & Strong\\
CMB+BAO+JLA & IDE2 & $-8.3$ & Very Strong \\
CMB+BAO+JLA+CC & IDE2 & $-8.9$ & Very Strong \\
CMB+R19 & IDE2 & $+1.6$ & Definite/Positive for IDE2\\
CMB+BAO+R19 & IDE2 & $-2.2$ & Definite/Positive\\
\hline 
CMB & IDE3 & $-2.1$ & Definite/Positive\\
CMB+BAO & IDE3 & $-7.6$ & Strong\\
CMB+BAO+JLA & IDE3 & $-8.8$ & Strong\\
CMB+BAO+JLA+CC & IDE3 & $-9.5$ & Strong\\ 
CMB+R19 & IDE3 & $+2.0$ & Definite/Positive for IDE3\\
CMB+BAO+R19 & IDE3 & $-1.1$ & Definite/Positive\\
\hline 
CMB & IDE4 & $-2.0$ & Definite/Positive\\
CMB+BAO & IDE4 & $-5.2$ & Definite/Positive\\
CMB+BAO+JLA & IDE4 & $-9.6$ & Strong\\
CMB+BAO+JLA+CC & IDE4 & $-9.7$ & Strong\\ 
CMB+R19 & IDE4 & $+0.9$ & Weak for IDE4\\
CMB+BAO+R19 & IDE4 & $-2.3$ & Definite/Positive\\
\hline

\hline\hline
\end{tabular}    
\caption{The   values of $\ln B_{ij}$, where $j$ stands for the 
reference model $\Lambda$CDM and $i$ for the IDE models. The negative sign indicates that 
the reference model is favored over the IDE models. } 
\label{tab:bayesian}                          
\end{center}    
\end{table*} 
for computing the evidences, since the code  directly accepts the MCMC chains of the 
analysis.   We refer to 
Ref.~\cite{Yang:2018qmz} for the discussions on Bayesian evidence analysis, and in
Table~\ref{tab:jeffreys} we provide the     revised Jeffreys scale by Kass and 
Raftery~\cite{Kass:1995loi}.

For all the examined scenarios we compute the values of $\ln B_{ij}$, which are 
summarized in Table 
\ref{tab:bayesian}. From this Table one can see that $\Lambda$CDM paradigm is most of 
the time preferred over the present IDE models, with the exception of the CMB+R19 
combination, where we see a weak/positive evidence for all  IDE models against 
$\Lambda$CDM.  This is expected since the 
number of free parameters of all IDE models is   eight, namely two more compared to the 
six parameters $\Lambda$CDM. 
                                                                                          
\section{Concluding remarks}
\label{sec-summary}

Interacting scenarios have attracted the interest of the literature, since they are 
efficient in alleviating the coincidence problem, and additionally they seem to alleviate 
the $H_0$ tension and $\sigma_8$ tensions. In the present work we 
investigated interacting scenarios which belong to a wider class, since they include a  
dynamical dark energy component whose equation of state  follows various  one-parameter 
parametrizations. In particular, our focus was to see if a non-zero interaction is 
favoured, and if the $H_0$ tension is still alleviated.

We considered  a   well known interaction in the literature  of the form $Q = 3 H 
\xi (1+w_x) \rho_x$, and we took
the dark energy 
equation-of-state parameter $w_x$ to have the 
expressions: 
$w_x(a)=w_0a[1-\log(a)]$ (Model IDE1), 
$w_x = w_0 a \exp(1-a)$ (Model IDE2), 
$w_x(a)=w_0a[1+\sin(1-a)]$ (Model IDE3), and 
$w_x(a)=w_0a[1+\arcsin(1-a)]$ (Model IDE4). Additionally, we  used the  latest 
observational data from 
CMB, JLA, BAO, Hubble parameter measurements from CC, and a gaussian prior on $H_0$  
labeled as R19  from SH0ES~\cite{Riess:2019cxk}. 

Our analysis shows that the coupling strength for all   interacting scenarios is 
quite small, and 
thus the models are consistent  with the non-interacting $w_x$-cosmology. In 
particular, all   scenarios are in agreement with $\xi = 0$ within 2$\sigma$, 
but an indication for $\xi$ greater than zero appears at $1\sigma$ when JLA and JLA+CC 
are 
added to CMB+BAO, or when R19 is added to both  CMB and CMB+BAO in the IDE1 and IDE2 scenarios. 

Concerning the current value of the dark energy equation of state $w_0$, for all 
interacting scenarios and for all  combination of datasets it always lies in the phantom 
regime at more than two/three standard deviations.  Moreover, we find a robust 
anti-correlation between $w_0$ and $H_0$.

However, the most striking feature, and one of the main results of the present work, is 
that  for all   interacting models, independently of 
the combination of datasets considered,  the estimated values of the Hubble parameter 
$H_0$ are greater compared to 
the $\Lambda$CDM-based Planck's estimation \cite{Ade:2015xua} and close to the local 
measurements of $H_0$ from Riess et al. 2016 \cite{Riess:2016jrr}, Riess et al. 2018 
\cite{R18} and Riess et al. 2019 \cite{Riess:2019cxk}. This is triggered by the 
aforementioned anti-correlation between $w_0$ and $H_0$ and the strongly phantom values 
we 
obtain for $w_0$. 
The alleviation of $H_0$ tension is independent of the interaction model due to the 
absence of correlation between 
$\xi$ and $H_0$, as shown in the two dimensional joint contours obtained for all   
observational datasets.

In summary, the extended interacting scenarios that include  dark energy sectors with a 
dynamical equation of state with 
only one free parameter, are very efficient in alleviating the    $H_0$ tension.
 
\vspace{0.2cm}
\begin{acknowledgments}
The authors thank the referee for some useful comments. SP has been supported by the Mathematical Research Impact-Centric Support Scheme 
(MATRICS), File No. MTR/2018/000940, given by the Science and Engineering Research Board 
(SERB), Govt. of India, as well as  by the Faculty Research and Professional Development 
Fund (FRPDF) Scheme of Presidency University, Kolkata, India. WY was  supported by the 
National Natural Science Foundation of China under Grants 
No. 11705079 and No. 11647153. 
EDV was supported from the European Research Council in the form of a Consolidator Grant 
with number 681431. SC acknowledges the Mathematical Research Impact Centric Support 
(MATRICS), File No. MTR/2017/000407, by the 
Science and Engineering Research Board (SERB), Government of India. This article is based 
upon work from CANTATA COST (European Cooperation in Science and Technology) action 
CA15117, EU Framework Programme Horizon 2020.
\end{acknowledgments}


\begin{thebibliography}{}

\bibitem{Copeland:2006wr} 
  E.~J.~Copeland, M.~Sami and S.~Tsujikawa,
  {\it Dynamics of dark energy,}
  Int.\ J.\ Mod.\ Phys.\ D {\bf 15}, 1753 (2006)
  [hep-th/0603057].
  
  \bibitem{Capozziello:2011et}
S.~Capozziello and M.~De Laurentis,
 {\it{Extended Theories of Gravity}},
Phys.\ Rept.\ {\bf 509}, 167 (2011)
[arXiv:1108.6266 [gr-qc]].


  \bibitem{Cai:2015emx} 
  Y.~F.~Cai, S.~Capozziello, M.~De Laurentis and E.~N.~Saridakis,
  {\it f(T) teleparallel gravity and cosmology,}
  Rept.\ Prog.\ Phys.\  {\bf 79}, no. 10, 106901 (2016)
  [arXiv:1511.07586 [gr-qc]].
 

\bibitem{Nojiri:2017ncd} 
  S.~Nojiri, S.~D.~Odintsov and V.~K.~Oikonomou,
  {\it Modified Gravity Theories on a Nutshell: Inflation, Bounce and Late-time 
Evolution,}
  Phys.\ Rept.\  {\bf 692}, 1 (2017)
  [arXiv:1705.11098 [gr-qc]].

  \bibitem{Wetterich-ide1} C. Wetterich, {\it The cosmon model for an asymptotically 
vanishing time-dependent cosmological ``constant'',} Astron. Astrophys. \textbf{301}, 321 
(1995) [arXiv:hep-th/9408025].


\bibitem{Amendola-ide1} L. Amendola, {\it Coupled Quintessence,}
Phys. Rev. D \textbf{62}, 043511 (2000) 
[arXiv:astro-ph/9908023].

\bibitem{Amendola-ide2} L. Amendola and C. Quercellini, {\it Tracking and coupled dark 
energy as seen by WMAP,} 
Phys. Rev. D \textbf{68}, 023514 (2003) 
[arXiv:astro-ph/0303228].

\bibitem{Cai:2004dk} 
  R.~G.~Cai and A.~Wang,
  {\it Cosmology with interaction between phantom dark energy and dark matter and the 
coincidence problem,}
  JCAP {\bf 0503}, 002 (2005)
  [hep-th/0411025].

\bibitem{Pavon:2005yx} 
D.~Pav\'{o}n and W.~Zimdahl, {\it Holographic dark energy and cosmic coincidence,}
Phys.\ Lett.\ B {\bf 628}, 206 (2005)
[arXiv:gr-qc/0505020]. 


\bibitem{delCampo:2008sr} 
S.~del Campo, R.~Herrera and D.~Pav\'{o}n, {\it Toward a solution of the coincidence 
problem,}
Phys.\ Rev.\ D {\bf 78}, 021302 (2008)
[arXiv:0806.2116 [astro-ph]].

\bibitem{delCampo:2008jx} 
S.~del Campo, R.~Herrera and D.~Pav\'{o}n, 
{\it Interacting models may be key to solve the cosmic coincidence problem,} J. Cosmol. 
Astropart. Phys. {\bf 0901}, 020 (2009)
[arXiv:0812.2210 [gr-qc]].

 \bibitem{Barrow:2006hia} 
  J.~D.~Barrow and T.~Clifton,
  {\it Cosmologies with energy exchange,}
  Phys.\ Rev.\ D {\bf 73}, 103520 (2006)
  [gr-qc/0604063].
  
  \bibitem{Amendola:2006dg}
  L.~Amendola, G.~Camargo Campos and R.~Rosenfeld,
  {\it Consequences of dark matter-dark energy interaction on cosmological parameters 
derived from SNIa data,}
  Phys.\ Rev.\ D {\bf 75}, 083506 (2007)
  [astro-ph/0610806].

  
\bibitem{He:2008tn}
  J.~H.~He and B.~Wang,
  {\it Effects of the interaction between dark energy and dark matter on cosmological 
parameters,}
  JCAP {\bf 0806}, 010 (2008)
  [arXiv:0801.4233 [astro-ph]].
  
  \bibitem{Chen:2008ft} 
  X.~m.~Chen, Y.~g.~Gong and E.~N.~Saridakis,
  {\it Phase-space analysis of interacting phantom cosmology,}
  JCAP {\bf 0904}, 001 (2009)
  [arXiv:0812.1117 [gr-qc]].
  
  
 
\bibitem{Basilakos:2008ae} 
  S.~Basilakos and M.~Plionis,
  {\it Is the Interacting Dark Matter Scenario an Alternative to Dark Energy ?,}
  Astron.\ Astrophys.\  {\bf 507}, 47 (2009)
  [arXiv:0807.4590 [astro-ph]].
  

  
  
  
\bibitem{Gavela:2009cy} 
  M.~B.~Gavela, D.~Hernandez, L.~Lopez Honorez, O.~Mena and S.~Rigolin,
  {\it Dark coupling,}
  JCAP {\bf 0907}, 034 (2009)
  [arXiv:0901.1611 [astro-ph.CO]].
  
  
  \bibitem{Sadjadi:2009sp} 
  H.~M.~Sadjadi,
  {\it $w(d)= -1$ in interacting quintessence model,}
  Eur.\ Phys.\ J.\ C {\bf 66}, 445 (2010)
  [arXiv:0904.1349 [gr-qc]].
  
\bibitem{Jamil:2009eb} 
  M.~Jamil, E.~N.~Saridakis and M.~R.~Setare,
 {\it  Thermodynamics of dark energy interacting with dark matter and radiation},
  Phys.\ Rev.\ D {\bf 81}, 023007 (2010)
  [arXiv:0910.0822 [hep-th]].
  
  
\bibitem{Chen:2011cy} 
  X.~m.~Chen, Y.~Gong, E.~N.~Saridakis and Y.~Gong,
   {\it Time-dependent interacting dark energy and transient acceleration},
  Int.\ J.\ Theor.\ Phys.\  {\bf 53}, 469 (2014)
  [arXiv:1111.6743 [astro-ph.CO]].
  
  
   \bibitem{Pan:2013rha} 
  S.~Pan and S.~Chakraborty,
  {\it Will there be again a transition from acceleration to deceleration in course of 
the 
dark energy evolution of the universe?,}
  Eur.\ Phys.\ J.\ C {\bf 73}, 2575 (2013)
  [arXiv:1303.5602 [gr-qc]].
  
\bibitem{Yang:2014vza}
  W.~Yang and L.~Xu,
  {\it Testing coupled dark energy with large scale structure observation,}
  JCAP {\bf 1408}, 034 (2014)
  [arXiv:1401.5177 [astro-ph.CO]].
  
  \bibitem{Yang:2014gza}  
  W.~Yang and L.~Xu,
  {\it Cosmological constraints on interacting dark energy with redshift-space distortion 
after Planck data,}
  Phys.\ Rev.\ D {\bf 89}, no.8,  083517 (2014)
  [arXiv:1401.1286 [astro-ph.CO]].
  
  \bibitem{Nunes:2014qoa} 
  R.~C.~Nunes and E.~M.~Barboza,
  {\it Dark matter-dark energy interaction for a time-dependent EoS parameter,}
  Gen.\ Rel.\ Grav.\  {\bf 46}, 1820 (2014)
  [arXiv:1404.1620 [astro-ph.CO]].
  
\bibitem{Faraoni:2014vra} 
  V.~Faraoni, J.~B.~Dent and E.~N.~Saridakis,
  {\it Covariantizing the interaction between dark energy and dark matter,}
  Phys.\ Rev.\ D {\bf 90}, no. 6, 063510 (2014)
  [arXiv:1405.7288 [gr-qc]].
  
  \bibitem{Salvatelli:2014zta} 
  V.~Salvatelli, N.~Said, M.~Bruni, A.~Melchiorri and D.~Wands,
  {\it Indications of a late-time interaction in the dark sector,}
  Phys.\ Rev.\ Lett.\  {\bf 113}, no. 18, 181301 (2014)
  [arXiv:1406.7297 [astro-ph.CO]].
  
  \bibitem{Yang:2014hea} 
  W.~Yang and L.~Xu,
  {\it Coupled dark energy with perturbed Hubble expansion rate,}
  Phys.\ Rev.\ D {\bf 90}, no. 8, 083532 (2014)
  [arXiv:1409.5533 [astro-ph.CO]].
  
  \bibitem{Pan:2012ki}
  S.~Pan, S.~Bhattacharya and S.~Chakraborty,
  {\it An analytic model for interacting dark energy and its observational constraints,}
  Mon.\ Not.\ Roy.\ Astron.\ Soc.\  {\bf 452}, no.3,  3038 (2015)
  [arXiv:1210.0396 [gr-qc]].
  
  \bibitem{Li:2015vla} 
  Y.~H.~Li, J.~F.~Zhang and X.~Zhang,
  {\it Testing models of vacuum energy interacting with cold dark matter,}
  Phys.\ Rev.\ D {\bf 93}, no. 2, 023002 (2016)
  [arXiv:1506.06349 [astro-ph.CO]].
  
  \bibitem{Nunes:2016dlj} 
  R.~C.~Nunes, S.~Pan and E.~N.~Saridakis,
 {\it New constraints on interacting dark energy from cosmic chronometers},
  Phys.\ Rev.\ D {\bf 94}, no. 2, 023508 (2016)
  [arXiv:1605.01712 [astro-ph.CO]].
  
  \bibitem{Yang:2016evp}
  W.~Yang, H.~Li, Y.~Wu and J.~Lu,
  {\it Cosmological constraints on coupled dark energy,}
  JCAP {\bf 1610}, no.10,  007 (2016)
  [arXiv:1608.07039 [astro-ph.CO]].
 
  
  \bibitem{Kumar:2016zpg} 
  S.~Kumar and R.~C.~Nunes,
  {\it Probing the interaction between dark matter and dark energy in the presence of 
massive neutrinos,}
  Phys.\ Rev.\ D {\bf 94}, no. 12, 123511 (2016)
  [arXiv:1608.02454 [astro-ph.CO]].
 
  \bibitem{Pan:2016ngu} 
  S.~Pan and G.~S.~Sharov,
  {\it A model with interaction of dark components and recent observational data,}
  Mon.\ Not.\ Roy.\ Astron.\ Soc.\  {\bf 472}, no. 4, 4736 (2017)
  [arXiv:1609.02287 [gr-qc]].
  
  \bibitem{Mukherjee:2016shl} 
  A.~Mukherjee and N.~Banerjee,
  {\it In search of the dark matter dark energy interaction: a kinematic approach,}
  Class.\ Quant.\ Grav.\  {\bf 34}, no. 3, 035016 (2017)
  [arXiv:1610.04419 [astro-ph.CO]].
  
  \bibitem{Erdem:2016hqw} 
  R.~Erdem,
  {\it Is it possible to obtain cosmic accelerated expansion through energy transfer 
between different energy densities?,}
  Phys.\ Dark Univ.\  {\bf 15}, 57 (2017)
  [arXiv:1612.04864 [gr-qc]].
  
  \bibitem{Sharov:2017iue} 
  G.~S.~Sharov, S.~Bhattacharya, S.~Pan, R.~C.~Nunes and S.~Chakraborty,
  {\it A new interacting two fluid model and its consequences,}
  Mon.\ Not.\ Roy.\ Astron.\ Soc.\  {\bf 466}, no. 3, 3497 (2017)
  [arXiv:1701.00780 [gr-qc]].
  
  \bibitem{Shahalam:2017fqt} 
  M.~Shahalam, S.~D.~Pathak, S.~Li, R.~Myrzakulov and A.~Wang,
  {\it Dynamics of coupled phantom and tachyon fields,}
  Eur.\ Phys.\ J.\ C {\bf 77}, no. 10, 686 (2017)
  [arXiv:1702.04720 [gr-qc]].
  
  
\bibitem{Guo:2017hea} 
  R.~Y.~Guo, Y.~H.~Li, J.~F.~Zhang and X.~Zhang,
  {\it Weighing neutrinos in the scenario of vacuum energy interacting with cold dark 
matter: application of the parameterized post-Friedmann approach,}
  JCAP {\bf 1705}, no. 05, 040 (2017)
  [arXiv:1702.04189 [astro-ph.CO]].
  
  \bibitem{Cai:2017yww} 
  R.~G.~Cai, N.~Tamanini and T.~Yang,
  {\it Reconstructing the dark sector interaction with LISA,}
  JCAP {\bf 1705}, no. 05, 031 (2017)
  [arXiv:1703.07323 [astro-ph.CO]].
  
  \bibitem{Yang:2017yme} 
  W.~Yang, N.~Banerjee and S.~Pan,
  {\it Constraining a dark matter and dark energy interaction scenario with a dynamical 
equation of state,}
  Phys.\ Rev.\ D {\bf 95}, no. 12, 123527 (2017)
  [arXiv:1705.09278 [astro-ph.CO]].
  
  \bibitem{Santos:2017bqm} 
  L.~Santos, W.~Zhao, E.~G.~M.~Ferreira and J.~Quintin,
  {\it Constraining interacting dark energy with CMB and BAO future surveys,}
  Phys.\ Rev.\ D {\bf 96}, no. 10, 103529 (2017)
  [arXiv:1707.06827 [astro-ph.CO]].
  
  \bibitem{Yang:2017ccc} 
  W.~Yang, S.~Pan and D.~F.~Mota,
  {\it Novel approach toward the large-scale stable interacting dark-energy models and 
their astronomical bounds,}
  Phys.\ Rev.\ D {\bf 96}, no. 12, 123508 (2017)
  [arXiv:1709.00006 [astro-ph.CO]].
  
  \bibitem{vandeBruck:2017idm} 
  C.~Van De Bruck and J.~Mifsud,
  {\it Searching for dark matter - dark energy interactions: going beyond the conformal 
case,}
  Phys.\ Rev.\ D {\bf 97}, no. 2, 023506 (2018)
  [arXiv:1709.04882 [astro-ph.CO]].
  
  \bibitem{Pan:2017ent} 
  S.~Pan, A.~Mukherjee and N.~Banerjee,
  {\it Astronomical bounds on a cosmological model allowing a general interaction in the 
dark sector,}
  Mon.\ Not.\ Roy.\ Astron.\ Soc.\  {\bf 477}, no. 1, 1189 (2018)
  [arXiv:1710.03725 [astro-ph.CO]].
  
  \bibitem{Xu:2017rfo} 
  X.~Xu, Y.~Z.~Ma and A.~Weltman,
  {\it Constraining the interaction between dark sectors with future HI intensity mapping 
observations,}
  Phys.\ Rev.\ D {\bf 97}, no. 8, 083504 (2018)
  [arXiv:1710.03643 [astro-ph.CO]].
  
  \bibitem{Yang:2018xlt} 
  W.~Yang, S.~Pan, R.~Herrera and S.~Chakraborty,
  {\it Large-scale (in) stability analysis of an exactly solved coupled dark-energy 
model,}
  Phys.\ Rev.\ D {\bf 98}, no. 4, 043517 (2018)
  [arXiv:1808.01669 [gr-qc]].
  
   \bibitem{Yang:2018ubt} 
  W.~Yang, S.~Pan, L.~Xu and D.~F.~Mota,
  {\it Effects of anisotropic stress in interacting dark matter-dark energy scenarios,}
  Mon.\ Not.\ Roy.\ Astron.\ Soc.\  {\bf 482}, no. 2, 1858 (2019)
  [arXiv:1804.08455 [astro-ph.CO]].
  
  \bibitem{Yang:2018pej} 
  W.~Yang, S.~Pan and A.~Paliathanasis,
  {\it Cosmological constraints on an exponential interaction in the dark sector,}
  Mon.\ Not.\ Roy.\ Astron.\ Soc.\  {\bf 482}, no. 1, 1007 (2019)
  [arXiv:1804.08558 [gr-qc]].
  
  \bibitem{Cardenas:2018qcg} 
  D.~Grandon and V.~H.~Cardenas,
  {\it Exploring evidence of interaction between dark energy and dark matter,}
  Gen.\ Rel.\ Grav.\  {\bf 51}, no. 42 (2019)
  [arXiv:1804.03296 [astro-ph.CO]].

\bibitem{Odintsov:2018awm} 
  S.~D.~Odintsov and V.~K.~Oikonomou,
  {\it Study of finite-time singularities of loop quantum cosmology interacting 
multifluids,}
  Phys.\ Rev.\ D {\bf 97}, no. 12, 124042 (2018)
  [arXiv:1806.01588 [gr-qc]].

\bibitem{vonMarttens:2018iav} 
  R.~von Marttens, L.~Casarini, D.~F.~Mota and W.~Zimdahl,
  {\it Cosmological constraints on parametrized interacting dark energy,}
  Phys.\ Dark Univ.\  {\bf 23}, 100248 (2019)
  [arXiv:1807.11380 [astro-ph.CO]].
  
  \bibitem{Yang:2018qec} 
  W.~Yang, N.~Banerjee, A.~Paliathanasis and S.~Pan,
  {\it Reconstructing the dark matter and dark energy interaction scenarios from  
observations,}
  arXiv:1812.06854 [astro-ph.CO].
  
  \bibitem{Bonici:2018qli} 
  M.~Bonici and N.~Maggiore,
  {\it Constraints on interacting dynamical dark energy from the cosmological equation of 
state,}
  arXiv:1812.11176 [gr-qc].
  
  \bibitem{Asghari:2019qld} 
  M.~Asghari, J.~B.~Jim\'{e}nez, S.~Khosravi and D.~F.~Mota,
  {\it On structure formation from a small-scales-interacting dark sector,}
  JCAP {\bf 1904}, no. 04, 042 (2019)
  [arXiv:1902.05532 [astro-ph.CO]].

\bibitem{Paliathanasis:2019hbi} 
  A.~Paliathanasis, S.~Pan and W.~Yang,
  {\it Dynamics of nonlinear interacting dark energy models,}
  arXiv:1903.02370 [gr-qc].
 
  \bibitem{Pan:2019jqh} 
  S.~Pan, W.~Yang, C.~Singha and E.~N.~Saridakis,
  {\it Observational constraints on sign-changeable interaction models and alleviation of 
the $H_0$ tension,}
  arXiv:1903.10969 [astro-ph.CO].
  
  \bibitem{Feng:2019mym}
  L.~Feng, H.~L.~Li, J.~F.~Zhang and X.~Zhang,
  {\it Exploring neutrino mass and mass hierarchy in interacting dark energy models,}
  arXiv:1903.08848 [astro-ph.CO].
  
  
  \bibitem{Li:2019loh} 
  C.~Li, X.~Ren, M.~Khurshudyan and Y.~F.~Cai,
  {\it Implications of the possible 21-cm line excess at cosmic dawn on dynamics of 
interacting dark energy,}
  arXiv:1904.02458 [astro-ph.CO].
  
  \bibitem{Yang:2019bpr} 
  W.~Yang, S.~Pan, E.~Di Valentino, B.~Wang and A.~Wang,
  {\it Forecasting Interacting Vacuum-Energy Models using Gravitational Waves,}
  arXiv:1904.11980 [astro-ph.CO].
  
 
 
  \bibitem{Yang:2019vni} 
  W.~Yang, S.~Vagnozzi, E.~Di Valentino, R.~C.~Nunes, S.~Pan and D.~F.~Mota, {\it 
Listening to the sound of dark sector interactions with gravitational wave standard 
sirens,} to appear in JCAP,
  arXiv:1905.08286 [astro-ph.CO].
  
  \bibitem{Oikonomou:2019nmm} 
  V.~K.~Oikonomou,
  {\it Generalized Logarithmic Equation of State in Classical and Loop Quantum Cosmology 
Dark Energy-Dark Matter Coupled Systems,}
  arXiv:1907.02600 [gr-qc].
  
  \bibitem{Bolotin:2013jpa} 
  Y.~L.~Bolotin, A.~Kostenko, O.~A.~Lemets and D.~A.~Yerokhin,
  {\it Cosmological Evolution With Interaction Between Dark Energy And Dark Matter,}
  Int.\ J.\ Mod.\ Phys.\ D {\bf 24}, no. 03, 1530007 (2015)
  [arXiv:1310.0085 [astro-ph.CO]].
  
  \bibitem{Wang:2016lxa}
  B.~Wang, E.~Abdalla, F.~Atrio-Barandela and D.~Pav\'{o}n,
  {\it Dark Matter and Dark Energy Interactions: Theoretical Challenges, Cosmological 
Implications and Observational Signatures,}
  Rept.\ Prog.\ Phys.\  {\bf 79} (2016) no.9,  096901
  [arXiv:1603.08299 [astro-ph.CO]].
  
  
  \bibitem{Wang:2005jx} 
  B.~Wang, Y.~g.~Gong and E.~Abdalla,
  {\it Transition of the dark energy equation of state in an interacting holographic dark 
energy model,}
  Phys.\ Lett.\ B {\bf 624}, 141 (2005)
  [hep-th/0506069].
  
  
 \bibitem{Sadjadi:2006qb} 
  H.~M.~Sadjadi and M.~Honardoost,
  {\it Thermodynamics second law and $omega = -1$ crossing(s) in interacting holographic 
dark energy model,}
  Phys.\ Lett.\ B {\bf 647}, 231 (2007)
  [gr-qc/0609076].

\bibitem{Pan:2014afa} 
  S.~Pan and S.~Chakraborty,
  {\it A cosmographic analysis of holographic dark energy models,}
  Int.\ J.\ Mod.\ Phys.\ D {\bf 23}, no. 11, 1450092 (2014)
  [arXiv:1410.8281 [gr-qc]].
  
  
 
 \bibitem{DiValentino:2015ola} 
  E.~Di Valentino, A.~Melchiorri and J.~Silk,
 {\it Beyond six parameters: extending $\Lambda$CDM,}
  Phys.\ Rev.\ D {\bf 92}, no. 12, 121302 (2015)
  [arXiv:1507.06646 [astro-ph.CO]].
  
\bibitem{DiValentino:2016hlg} 
  E.~Di Valentino, A.~Melchiorri and J.~Silk,
 {\it Reconciling Planck with the local value of $H_0$ in extended parameter space,}
  Phys.\ Lett.\ B {\bf 761}, 242 (2016)
  [arXiv:1606.00634 [astro-ph.CO]].
  
  \bibitem{Kumar:2017dnp} 
  S.~Kumar and R.~C.~Nunes,
  {\it Echo of interactions in the dark sector,} 
  Phys.\ Rev.\ D {\bf 96}, no. 10, 103511 (2017)
  [arXiv:1702.02143 [astro-ph.CO]].
  
  \bibitem{DiValentino:2017iww} 
  E.~Di Valentino, A.~Melchiorri and O.~Mena,
  {\it Can interacting dark energy solve the $H_0$ tension?,}
  Phys.\ Rev.\ D {\bf 96}, no. 4, 043503 (2017)
  [arXiv:1704.08342 [astro-ph.CO]].
  
  
  \bibitem{DiValentino:2017zyq} 
  E.~Di Valentino, A.~Melchiorri, E.~V.~Linder and J.~Silk,
 {\it Constraining Dark Energy Dynamics in Extended Parameter Space,}
  Phys.\ Rev.\ D {\bf 96}, no. 2, 023523 (2017)
  [arXiv:1704.00762 [astro-ph.CO]].
  
  \bibitem{Renk:2017rzu} 
  J.~Renk, M.~Zumalac\'{a}rregui, F.~Montanari and A.~Barreira,
  {\it Galileon gravity in light of ISW, CMB, BAO and H$_0$ data,}
  JCAP {\bf 1710}, no. 10, 020 (2017)
  [arXiv:1707.02263 [astro-ph.CO]].
  
\bibitem{DiValentino:2017gzb} 
  E.~Di Valentino,
{\it Crack in the cosmological paradigm,}
  Nat.\ Astron.\  {\bf 1}, no. 9, 569 (2017)
  [arXiv:1709.04046 [physics.pop-ph]].

\bibitem{DiValentino:2017oaw} 
  E.~Di Valentino, C.~B{\o}ehm, E.~Hivon and F.~R.~Bouchet,
  {\it Reducing the $H_0$ and $\sigma_8$ tensions with Dark Matter-neutrino interactions,}
  Phys.\ Rev.\ D {\bf 97}, no. 4, 043513 (2018)
  [arXiv:1710.02559 [astro-ph.CO]].
  

\bibitem{Fernandez-Arenas:2017isq} 
  D.~Fernandez Arenas {\it et al.},
  {\it An independent determination of the local Hubble constant},
  Mon.\ Not.\ Roy.\ Astron.\ Soc.\  {\bf 474}, no. 1, 1250 (2018)
  [arXiv:1710.05951 [astro-ph.CO]].
  

  
\bibitem{DiValentino:2017rcr} 
  E.~Di Valentino, E.~V.~Linder and A.~Melchiorri,
  {\it Vacuum phase transition solves the $H_0$ tension,}
  Phys.\ Rev.\ D {\bf 97}, no. 4, 043528 (2018)
  [arXiv:1710.02153 [astro-ph.CO]].
 

 \bibitem{Khosravi:2017hfi} 
  N.~Khosravi, S.~Baghram, N.~Afshordi and N.~Altamirano,
  {\it $H_0$ tension as a hint for a transition in gravitational theory,}
  Phys.\ Rev.\ D {\bf 99}, no. 10, 103526 (2019)
  [arXiv:1710.09366 [astro-ph.CO]].
  
  \bibitem{Mortsell:2018mfj} 
  E.~M\"{o}rtsell and S.~Dhawan,
  {\it Does the Hubble constant tension call for new physics?,}
  JCAP {\bf 1809}, no. 09, 025 (2018)
  [arXiv:1801.07260 [astro-ph.CO]].
  
  \bibitem{Yang:2018euj} 
  W.~Yang, S.~Pan, E.~Di Valentino, R.~C.~Nunes, S.~Vagnozzi and D.~F.~Mota,
  {\it Tale of stable interacting dark energy, observational signatures, and the $H_0$ 
tension,}
  JCAP {\bf 1809}, no. 09, 019 (2018)
  [arXiv:1805.08252 [astro-ph.CO]].
  
  \bibitem{DEramo:2018vss} 
  F.~D'Eramo, R.~Z.~Ferreira, A.~Notari and J.~L.~Bernal,
  {\it Hot Axions and the $H_0$ tension,}
  JCAP {\bf 1811}, no. 11, 014 (2018)
  [arXiv:1808.07430 [hep-ph]].
  
\bibitem{Yang:2018uae} 
  W.~Yang, A.~Mukherjee, E.~Di Valentino and S.~Pan,
  {\it Interacting dark energy with time varying equation of state and the $H_0$ tension,}
  Phys.\ Rev.\ D {\bf 98}, no. 12, 123527 (2018)
  [arXiv:1809.06883 [astro-ph.CO]].
  
  \bibitem{Guo:2018ans} 
  R.~Y.~Guo, J.~F.~Zhang and X.~Zhang,
  {\it Can the $H_0$ tension be resolved in extensions to $\Lambda$CDM cosmology?,}
  JCAP {\bf 1902}, 054 (2019)
  [arXiv:1809.02340 [astro-ph.CO]].
  
  \bibitem{Yang:2018qmz}
  W.~Yang, S.~Pan, E.~Di Valentino, E.~N.~Saridakis and S.~Chakraborty, {\it 
Observational 
constraints on one-parameter dynamical dark-energy parametrizations and the $H_0$ 
tension,}  Phys. Rev. D {\bf 99} no.4, 043543 (2019),
  arXiv:1810.05141 [astro-ph.CO].
  
  \bibitem{Poulin:2018cxd} 
  V.~Poulin, T.~L.~Smith, T.~Karwal and M.~Kamionkowski,
 {\it Early Dark Energy Can Resolve The Hubble Tension,}
  Phys.\ Rev.\ Lett.\  {\bf 122}, no. 22, 221301 (2019)
  [arXiv:1811.04083 [astro-ph.CO]].
  
  \bibitem{Zhang:2018air} 
  X.~Zhang and Q.~G.~Huang,
  {\it Constraints on $H_0$ from WMAP and baryon acoustic osillation measurements,}
  arXiv:1812.01877 [astro-ph.CO].
 
 \bibitem{Kreisch:2019yzn} 
  C.~D.~Kreisch, F.~Y.~Cyr-Racine and O.~Dor\'{e},
  {\it The Neutrino Puzzle: Anomalies, Interactions, and Cosmological Tensions,}
  arXiv:1902.00534 [astro-ph.CO].
 
 \bibitem{Martinelli:2019dau} 
  M.~Martinelli, N.~B.~Hogg, S.~Peirone, M.~Bruni and D.~Wands,
 {\it Constraints on the interacting vacuum - geodesic CDM scenario,}
  arXiv:1902.10694 [astro-ph.CO].
  
  \bibitem{Pandey:2019plg} 
  K.~L.~Pandey, T.~Karwal and S.~Das,
  {\it Alleviating the $H_0$ and $\sigma_8$ anomalies with a decaying dark matter model,}
  arXiv:1902.10636 [astro-ph.CO].
  
  \bibitem{Vattis:2019efj} 
  K.~Vattis, S.~M.~Koushiappas and A.~Loeb,
 {\it Dark matter decaying in the late Universe can relieve the H0 tension,}
  Phys.\ Rev.\ D {\bf 99}, no. 12, 121302 (2019)
  [arXiv:1903.06220 [astro-ph.CO]].
  
  \bibitem{Kumar:2019wfs} 
  S.~Kumar, R.~C.~Nunes and S.~K.~Yadav,
  {\it Dark sector interaction: a remedy of the tensions between CMB and LSS data,}
  Eur.\ Phys.\ J.\ C {\bf 79}, no. 7, 576 (2019)
  [arXiv:1903.04865 [astro-ph.CO]].
 
\bibitem{Agrawal:2019lmo} 
  P.~Agrawal, F.~Y.~Cyr-Racine, D.~Pinner and L.~Randall,
  {\it Rock 'n' Roll Solutions to the Hubble Tension,}
  arXiv:1904.01016 [astro-ph.CO].
  
  
   \bibitem{Yang:2019qza} 
  W.~Yang, S.~Pan, E.~Di Valentino, A.~Paliathanasis and J.~Lu,
 {\it Challenging bulk viscous unified scenarios with cosmological observations,}
  arXiv:1906.04162 [astro-ph.CO].
 
\bibitem{Yang:2019uzo} 
  W.~Yang, O.~Mena, S.~Pan and E.~Di Valentino,
  {\it Dark sectors with dynamical coupling,}
  arXiv:1906.11697 [astro-ph.CO].
  

\bibitem{DiValentino:2019exe} 
  E.~Di Valentino, R.~Z.~Ferreira, L.~Visinelli and U.~Danielsson,
  {\it Late time transitions in the quintessence field and the $H_0$ tension,}
  arXiv:1906.11255 [astro-ph.CO].
  

 \bibitem{Yang:2019nhz} 
  W.~Yang, S.~Pan, S.~Vagnozzi, E.~Di Valentino, D.~F.~Mota and S.~Capozziello,
  {\it Dawn of the dark: unified dark sectors and the EDGES Cosmic Dawn 21-cm signal,}
  arXiv:1907.05344 [astro-ph.CO]. 
  
  
 \bibitem{Pourtsidou:2016ico} 
  A.~Pourtsidou and T.~Tram,
  {\it Reconciling CMB and structure growth measurements with dark energy interactions,}
  Phys.\ Rev.\ D {\bf 94}, no. 4, 043518 (2016)
  [arXiv:1604.04222 [astro-ph.CO]].
  
  

\bibitem{An:2017crg} 
  R.~An, C.~Feng and B.~Wang,
  {\it Relieving the Tension between Weak Lensing and Cosmic Microwave Background with 
Interacting Dark Matter and Dark Energy Models,}
  JCAP {\bf 1802}, no. 02, 038 (2018)
  [arXiv:1711.06799 [astro-ph.CO]].
 
  
  
\bibitem{DiValentino:2018gcu} 
  E.~Di Valentino and S.~Bridle,
  {\it Exploring the Tension between Current Cosmic Microwave Background and Cosmic Shear 
Data,}
  Symmetry {\bf 10}, no. 11, 585 (2018).
  
    
 \bibitem{Yang:2017zjs} 
  W.~Yang, S.~Pan and J.~D.~Barrow,
  {\it Large-scale Stability and Astronomical Constraints for Coupled Dark-Energy Models,}
  Phys.\ Rev.\ D {\bf 97}, no. 4, 043529 (2018)
  [arXiv:1706.04953 [astro-ph.CO]].
  
 
\bibitem {Mukhanov} V. F. Mukhanov, H. A. Feldman and R. H. Brandenberger, 
{\it Theory of cosmological perturbations,}
Phys. Rept. \textbf{215}, 203 (1992).


\bibitem {Ma:1995ey}C.~P.~Ma and E.~Bertschinger,
{\it Cosmological perturbation theory in the synchronous and conformal Newtonian gauges,}
Astrophys.\ J.\ \textbf{455}, 7 (1995)
arXiv:astro-ph/9506072.


\bibitem{Malik:2008im} 
  K.~A.~Malik and D.~Wands,
  {\it Cosmological perturbations,}
  Phys.\ Rept.\  {\bf 475}, 1 (2009)
  [arXiv:0809.4944 [astro-ph]].
  
  
\bibitem{Majerotto:2009np} 
  E.~Majerotto, J.~Valiviita and R.~Maartens,
  {\it Adiabatic initial conditions for perturbations in interacting dark energy models,}
  Mon.\ Not.\ Roy.\ Astron.\ Soc.\  {\bf 402}, 2344 (2010)
  [arXiv:0907.4981 [astro-ph.CO]].
  
  
\bibitem{Valiviita:2008iv} 
  J.~V\"{a}liviita, E.~Majerotto and R.~Maartens,
  {\it Instability in interacting dark energy and dark matter fluids,}
  JCAP {\bf 0807}, 020 (2008)
  [arXiv:0804.0232 [astro-ph]].

 
 \bibitem{Clemson:2011an} 
  T.~Clemson, K.~Koyama, G.~B.~Zhao, R.~Maartens and J.~Valiviita,
  {\it Interacting Dark Energy -- constraints and degeneracies,}
  Phys.\ Rev.\ D {\bf 85}, 043007 (2012)
  [arXiv:1109.6234 [astro-ph.CO]].
  
  
 
  
\bibitem{Adam:2015rua} 
  R.~Adam {\it et al.} [Planck Collaboration],
  {\it Planck 2015 results. I. Overview of products and scientific results,}
  Astron.\ Astrophys.\  {\bf 594}, A1 (2016)
  [arXiv:1502.01582 [astro-ph.CO]].

\bibitem{Aghanim:2015xee} 
  N.~Aghanim {\it et al.} [Planck Collaboration],
  {\it Planck 2015 results. XI. CMB power spectra, likelihoods, and robustness of 
parameters,}
  Astron.\ Astrophys.\  {\bf 594}, A11 (2016)
  [arXiv:1507.02704 [astro-ph.CO]].
  
  \bibitem{Betoule:2014frx} 
  M.~Betoule {\it et al.} [SDSS Collaboration],
  {\it Improved cosmological constraints from a joint analysis of the SDSS-II and SNLS 
supernova samples,}
  Astron.\ Astrophys.\  {\bf 568}, A22 (2014)
  [arXiv:1401.4064 [astro-ph.CO]].
  
\bibitem{Beutler:2011hx} 
  F.~Beutler {\it et al.},
  {\it The 6dF Galaxy Survey: Baryon Acoustic Oscillations and the Local Hubble Constant,}
  Mon.\ Not.\ Roy.\ Astron.\ Soc.\  {\bf 416}, 3017 (2011)
  [arXiv:1106.3366 [astro-ph.CO]].
  
  \bibitem{Ross:2014qpa} 
  A.~J.~Ross, L.~Samushia, C.~Howlett, W.~J.~Percival, A.~Burden and M.~Manera, {\it The 
clustering of the SDSS DR7 main Galaxy sample – I. A 4 per cent distance measure at $z = 
0.15$,}
  Mon.\ Not.\ Roy.\ Astron.\ Soc.\  {\bf 449}, no. 1, 835 (2015)
  [arXiv:1409.3242 [astro-ph.CO]].

\bibitem{Gil-Marin:2015nqa} 
  H.~Gil-Mar\'{i}n {\it et al.},
  {\it The clustering of galaxies in the SDSS-III Baryon Oscillation Spectroscopic 
Survey: 
BAO measurement from the LOS-dependent power spectrum of DR12 BOSS galaxies,}
  Mon.\ Not.\ Roy.\ Astron.\ Soc.\  {\bf 460}, no. 4, 4210 (2016)
  [arXiv:1509.06373 [astro-ph.CO]].
  

\bibitem{Moresco:2016mzx} 
  M.~Moresco {\it et al.},
  {\it A 6\% measurement of the Hubble parameter at $z\sim0.45$: direct evidence of the 
epoch of cosmic re-acceleration,}
  JCAP {\bf 1605}, no. 05, 014 (2016)
  [arXiv:1601.01701 [astro-ph.CO]].
  
   \bibitem{Riess:2019cxk}
  A.~G.~Riess, S.~Casertano, W.~Yuan, L.~M.~Macri and D.~Scolnic,
  {\it Large Magellanic Cloud Cepheid Standards Provide a 1\% Foundation for the 
Determination of the Hubble Constant and Stronger Evidence for Physics Beyond LambdaCDM,}
  arXiv:1903.07603 [astro-ph.CO]. 
  
  
\bibitem{Lewis:2002ah} 
  A.~Lewis and S.~Bridle,
  {\it Cosmological parameters from CMB and other data: A Monte Carlo approach,}
  Phys.\ Rev.\ D {\bf 66}, 103511 (2002)
  [astro-ph/0205436].


\bibitem{Lewis:1999bs} 
  A.~Lewis, A.~Challinor and A.~Lasenby,
  {\it Efficient computation of CMB anisotropies in closed FRW models,}
  Astrophys.\ J.\  {\bf 538}, 473 (2000)
  [astro-ph/9911177].
  
    \bibitem{Aghanim:2018eyx} 
  N.~Aghanim {\it et al.} [Planck Collaboration],
  {\it Planck 2018 results. VI. Cosmological parameters,}
  [arXiv:1807.06209 [astro-ph.CO]].
  
  
   \bibitem{Ade:2015xua} 
  P.~A.~R.~Ade {\it et al.} [Planck Collaboration],
  {\it Planck 2015 results. XIII. Cosmological parameters,}
  Astron.\ Astrophys.\  {\bf 594}, A13 (2016)
  [arXiv:1502.01589 [astro-ph.CO]].
  
\bibitem{Riess:2016jrr} 
  A.~G.~Riess {\it et al.},
  {\it A 2.4\% Determination of the Local Value of the Hubble Constant,}
  Astrophys.\ J.\  {\bf 826}, no. 1, 56 (2016)
  [arXiv:1604.01424 [astro-ph.CO]].
  
 \bibitem{R18}
A.~G.~Riess {\it et al.}, {\it New Parallaxes of Galactic Cepheids from Spatially 
Scanning 
the Hubble Space Telescope: Implications for the Hubble Constant}, 
Astrophys.\ J.\  {\bf 855}, 136 (2018). 
  
\bibitem{Heavens:2017afc} 
  A.~Heavens, Y.~Fantaye, A.~Mootoovaloo, H.~Eggers, Z.~Hosenie, S.~Kroon and 
E.~Sellentin,
  {\it Marginal Likelihoods from Monte Carlo Markov Chains,}
  arXiv:1704.03472 [stat.CO].
 
  
 \bibitem{Heavens:2017hkr} 
  A.~Heavens, Y.~Fantaye, E.~Sellentin, H.~Eggers, Z.~Hosenie, S.~Kroon and 
A.~Mootoovaloo,
  {\it No evidence for extensions to the standard cosmological model,}
  Phys.\ Rev.\ Lett.\  {\bf 119}, no. 10, 101301 (2017)
  [arXiv:1704.03467 [astro-ph.CO]].

  
\bibitem{Kass:1995loi}
  R.~E.~Kass and A.~E.~Raftery,
  {\it Bayes Factors,}
  J.\ Am.\ Statist.\ Assoc.\  {\bf 90}, no.430,  773 (1995). 

 
  

  

\end{thebibliography}
\end{document}